\documentclass{emulateapj}

\usepackage{subfigure}
\usepackage{apjfonts}
\usepackage{aas_macros}

\begin{document}
\title{X-ray counterpart of gravitational waves due to binary neutron star mergers: \\ light curves, luminosity function, and event rate density}
\author{Hui Sun\altaffilmark{1,2}, Bing Zhang\altaffilmark{1,2,3}, He Gao\altaffilmark{4}}

\altaffiltext{1}{Department of Astronomy, School of Physics, Peking University, Beijing 100871, China; hsun$\_$astro@pku.edu.cn}
\altaffiltext{2}{Kavli Institute for Astronomy and Astrophysics, Peking University, Beijing 100871, China;}
\altaffiltext{3}{Department of Physics and Astronomy, University of Nevada, Las Vegas, NV 89154, USA, zhang@physics.unlv.edu}
\altaffiltext{4}{Department of Astronomy, Beijing Normal University, Beijing 100875, China; gaohe@bnu.edu.cn}

\begin{abstract}
\cite{zhang13} proposed a type of GRB-less X-ray transient associated with double neutron star (NS-NS) mergers under the conjecture of a rapidly-spinning magnetar merger product with the line of sight off the short GRB jet. We investigate possible light curves of these transients by considering different observer's viewing angles. We perform Monte Carlo simulations to calculate the peak luminosity function (LF) and event rate density of these X-ray transients. By considering that a fraction of massive neutron stars may be supra-massive and later collapse into black holes after spinning down, we investigate how the predicted LF depends on the equation of state (EoS) of the central object and the geometry of the system. In general, the LF can be fit by two log-normal distributions peaking around $10^{46.4}$ and $10^{49.6}$ $\rm erg\,s^{-1}$, corresponding to the trapped and free zones, respectively. For the majority of the EoS models, the current non-detection is consistent with having a free zone solid angle at most a few times of the solid angle of the short GRB jet. The event rate density of these X-ray transients is around a few tens of $\rm Gpc^{-3}yr^{-1}$ for luminosity above $10^{45}$ $\rm erg\,s^{-1}$. We predict that future X-ray telescopes (such as Einstein Probe) with sensitivity $\sim 10^{-11}$ $\rm erg\,s^{-1}\,cm^{-2}$ would detect as many as several tens of such transients per year per steradian. Within 200 Mpc, the aLIGO average range for NS-NS mergers, the estimated event rate of these transients is about 1 transient per year all sky.
\end{abstract}

\section{Introduction}

 The detection of GW150914 \citep{abb16a}, and the follow-up detections of GW 151226 and LVT 151012 \citep{abb16b} by Advanced LIGO marked the beginning of gravitational wave astronomy. Detecting electromagnetic (EM) signals accompanying GW events is of great astrophysical significance, which allows one to locate the host galaxy and to study detailed physics of compact star mergers. Besides the already detected BH-BH mergers, 
NS-NS and NS-BH mergers are highly expected to be observed in the near future \citep{belczynski16}. 

 Several EM counterparts of GW events due to NS-NS and NS-BH mergers have been discussed in the literature:\footnote{Before GW 150914, BH-BH mergers have not been considered to have EM counterparts. \cite{connaughton16} reported a putative $\gamma$-ray counterpart to GW 150914, which drives interest in producing EM signals from BH-BH systems through EM processes \citep{zhang16,LP16} or accretion \citep{loeb16,perna16,li16a}.} 
 
The leading counterpart is short-duration GRBs (sGRBs), which are believed to be produced via accretion of merging materials into the central BH formed during the merger \citep{paczynski86,eichler89,paczynski91,narayan92,rosswog03,rezzolla11}. Current short GRB observations are in general consistent with such a hypothesis \citep{gehrels05,bar05,berger13,li16b}. Short GRBs are followed by long-lasting broad-band afterglows \citep{kann11,fong15}, which are also good targets for the observations of the EM counterparts of GW events.

The second EM counterpart is the optical/IR signal powered by the radioactivity of the neutron-rich materials launched during the NS-NS or NS-BH mergers \citep{LP98,kulkarni05,metzger10,MB12,BK13,TH13}. Such kind of signal, named as ``macronova'' or ``kilonova'', has been claimed to be associated with some short GRBs \citep{tanvir13,berger13,yang15,jin15,gao15,gao16b,jin16}.

The third EM counterpart is the afterglow-like emission associated with the macronova materials when they interact with the ambient medium. Since the ejecta is non-relativistic, the emission is mostly in the radio band and is called ``radio flares'' \citep{NP11,piran13,gao13}. No such candidate signal has been reported.

All the above EM signals are based on the assumption that the post-merger product is a BH. This is certainly true for NS-BH mergers. However, for NS-NS mergers it is possible that the post-merger product is a NS rather than a BH. There are four possibilities: a BH, a differential-rotation-supported hypermassive NS that lasts for $\sim 100$ ms before collapsing to a BH, a rigid-rotation-supported supra-massive NS that can survive for a much longer time (e.g. tens of seconds to $>10^4$ seconds) before collapsing after the NS is spun down enough, and a stable NS that lasts forever. The outcome of the merger product depends on the total mass in the NS-NS binary system, and the unknown NS equation of state (EoS). In most NS-NS merger simulations, typical 1.4 $M_\odot$ NSs are adopted. These simulations usually give rise to a BH or a hyper-massive NS \citep[e.g.][]{rosswog03,rezzolla11}. However, the total mass in Galactic NS-NS binaries are relatively small, typically in the range of (2.5-2.7) $M_\odot$ \citep{kiziltan13,martinez15}. The existence of NSs with masses of at least 2 $M_{\odot}$ \citep{lattimer10} suggests that the NS EoS is stiff. If the maximum mass of a non-rotating NS, $M_{\rm TOV}$, is about or even larger than 2.2 $M_\odot$, a supra-massive NS would be likely a merger product at least for some NS-NS mergers, since the maximum mass of a rigid-rotation-supported NS can be more massive than $M_{\rm TOV}$ by about 20\% \citep{lasky14,lv15,breu16}. Numerical simulations show that under extreme parameters (small NS masses and stiff EoS), even a stable NS can survive the merger \citep{GP13}. Observations of the afterglow of short GRBs show the existence of X-ray flares, shallow decay phase, and extended emission following short GRBs \citep{dai06,FX06,metzger08}, which may point towards a magnetar central engine. A smoking-gun evidence is the existence of an extended X-ray plateau (typically lasting 100s of seconds) followed by a very steep decay as observed in a good fraction of short GRBs \citep{rowlinson10,rowlinson13,lv15}. Such plateaus likely mark the existence of a supra-massive NS after the merger which collapse to a BH hundreds of seconds later, giving rise to rapid decay \citep{zhang14}. Global Monte Carlo simulations suggest that the short GRB data are consistent with such a picture for some NS EoSs \citep{gao16,li16c}. 

Invoking such a millisecond magnetar as the NS-NS post-merger product, \cite{zhang13} proposed a fourth EM counterpart for GW sources in the X-ray band. Since a millisecond magnetar launches an essentially isotropic magnetospheric wind, the effect of the magnetar can be observed even if the observer misses the jet emission that produces the short GRB. \cite{zhang13} hypothesized that the magnetar wind dissipates internally and produces X-rays via synchrotron radiation mechanism, probably through collision-induced magnetic reconnection \citep{ZY11} or current instability \citep{LB03}. This is based on the observations of the so-called ``internal" X-ray plateaus as observed in the X-ray lightcurves of some GRBs \citep{troja07,rowlinson10,rowlinson13,lv15}. These plateaus show a very steep decay at the end, which is impossible to account for within the models that invoke external shock interactions.\footnote{Several models invoking external processes to account for these X-ray plateaus have been proposed in the literature \citep[e.g.][]{RK15,siegel16a,siegel16b}, but none can reproduce the steep decay in the data.} \cite{zhang13} suggested that a wide-field X-ray detector may detect an X-ray signal coincident with a NS-NS merger GW event to be detected by LIGO and other GW observatories in the future.

Having a stable or supra-massive millisecond magnetar as the NS-NS merger product also has important implications for other EM counterparts of GW events. In particular, the steady energy injection of the magnetar spindown energy into the neutron rich ejecta would have two effects. First, the macro-/kilo-novae receives extra energy from the magnetar so that they are no longer simply powered by radioactive decay and could be much brighter than ``kilonova'' \citep{yu13,MP14}. In some extreme cases, these events can be even observed in X-rays \cite{siegel16a,siegel16b}\footnote{ \cite{siegel16a,siegel16b} suggested that most magnetar-powered merger-nova peak in the soft X-ray band. However, their model did not consider $p dV$ cooling of the merger-nova ejecta (D. M. Siegel, 2016, private communication). With such an effect considered, only magnetars with extreme parameters (e.g. $B \sim 10^{16}$ G, and $P_0 \sim 1$ ms) can produce an X-ray merger-nova. For typical magnetar parameters, the merger-nova is much fainter in X-rays, as shown in this paper.}. Following \cite{yu13}, we call such transients ``merger-novae''. \cite{gao16b} identified an optical bump signature in several short GRB afterglow lightcurves, which can be well interpreted as from the merger-novae. Second, the external shock emission when this ejecta interact with the ambient medium would be also brighter, which may be detectable in all wavelengths other than in radio \citep{gao13}. 

In this paper, we study the X-ray signature proposed by \cite{zhang13} in more detail. We introduce the concepts of ``free zone'' and ``trapped zone" for X-ray photons and calculate possible X-ray lightcurves for different viewing angles and different magnetar parameters (Sect. 2). Through Monte Carlo simulations, we derive the luminosity function of these transients. Since these X-ray transients have not been firmly detected, we use the non-detection to constrain the solid angle of the free zone as well as the event rate density of these transients  (Sect. 3). The fraction of supra-massive NSs also depend on the EoS of the NS (or quark star, QS), so we also investigate how our predictions depend on EoS. By comparing with other X-ray transients, we make predictions of the detectability of these transients by future wide-field X-ray detectors, such as Einstein Probe \citep{yuan16}.

\section{Model \& Lightcurves}

\subsection{Geometry}

\begin{figure}[hbtp]
\centering
\includegraphics[width=0.5\textwidth]{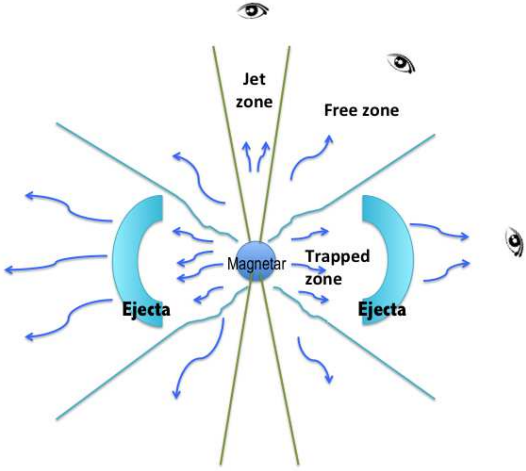} 
\caption{Cartoon figure that marks the jet, free and trapped region.}
\label{Fig:cartoon}
\end{figure}

We consider a NS-NS merger which leaves behind a stable or supra-massive millisecond magnetar behind. The post-merger geometric configuration may be delineated by a cartoon picture in Figure \ref{Fig:cartoon}. Since the open field angle of a millisecond magnetar is very wide, one can approximate a nearly isotropic pulsar wind. The X-rays produced by the internal dissipation of the magnetar wind are assumed to be emitted isotropically, whose luminosity tracks the dipole spin-down luminosity of the magnetar with a certain efficiency $\eta$. 
Numerical simulations of NS-NS mergers show that around $10^{-3}$ to $10^{-1}$ $M_{\odot}$ ejecta are ejected during the merger process \citep{freiburghaus99,rezzolla10,hotokezaka13,rosswog13}. These launched ejecta cover a significant solid angle. The X-ray photons can escape freely to the observer only when there is no ejecta in front. Otherwise, they are trapped by the ejecta and would first heat and accelerate the ejecta (along with the Poynting flux) and eventually escape when the ejecta becomes optically thin. 

Considering that such a system may also launch a relativistic jet in the direction perpendicular to the orbital plane \citep[e.g.][]{metzger08,ZD10,buc12}, one may define three zones (Figure \ref{Fig:cartoon}):
\begin{itemize}
 \item Jet zone: the direction where a short GRB can be detected. The X-rays from magnetar wind dissipation can be also observed, which powers the X-ray internal plateau as seen in a good fraction of short GRBs \citep{lv15};
 \item Free zone: the direction where no short GRB is observed but X-rays can still escape freely. In Fig. \ref{Fig:cartoon} this zone is marked as an annular ring around the jet, but in principle it can include the solid angle patches not covered by the ejecta in any direction. Since a millisecond magnetar wind is supposed to be essentially isotropic, throughout the paper, we assume that the X-ray luminosities in the free zone and in the jet zone are the same;
 \item Trapped zone: the direction where X-rays are initially trapped by the dynamical ejecta.
\end{itemize}

One has the sum of the solid angles
\begin{equation}
\Omega_{\rm jet} + \Omega_{\rm free} + \Omega_{\rm trapped} = 4\pi.
\end{equation}
Short GRB observations suggest a typical half opening angle of $10^{\rm o}$ (even though with a wide distribution) so that the average jet solid angle is $\left<\Omega_{\rm jet} \right>\sim (1/70) 4\pi$ \citep{berger14}. We then define a solid angle ratio parameter
  \begin{equation}
   k_{\rm \Omega}=\frac{\left<\Omega_{\rm free}\right>}{\left<\Omega_{\rm jet}\right>},
   \label{eq:k_Omega}
   \end{equation} 
with which the relative distributions among the three solid angles can be determined.

\subsection{Spin-down law and free zone emission}
The millisecond magnetar 
losses its rotation energy through both magnetic dipole radiation and gravitational wave (GW) radiation, with a total spin-down rate \citep[e.g.][]{shapiro83,usov92,ZM01,gao16,LG16}
\begin{equation}
\dot{E}=I\Omega \dot{\Omega}=-\frac{B_p^2R^6\Omega^4}{6c^3}-\frac{32GI^2 \epsilon ^2 \Omega^6}{5c^5}
\label{eq:Edot}
\end{equation}
where $\Omega=2\pi/P$ is the angular frequency and $\dot{\Omega}$ its time derivative, $I$ is the moment of inertia, $B_p$ is the dipolar field strength at the magnetic poles on the NS surface, $R$ is the radius of the NS, and $\epsilon$ is the ellipticity of the NS. The second term describes the GW radiation energy loss term, which describes the direct energy loss rate from the system, which does not affect the evolution of the remnant. The first term is the magnetic dipole radiation or magnetar wind spin-down term. In the following, we define the negative of this term as the dipole spin-down luminosity
\begin{equation}
L_{\rm sd}(t)=\frac{B_p^2 R^6 \Omega^4(t)}{6c^3},
\label{eqs:lsd}
\end{equation}
which is directly related to the total energy power of the EM counterparts in the system. Dropping out one of the two terms in the right hand side of Eq.(\ref{eq:Edot}), one can define two characteristic spin-down time scales \citep{ZM01}
\begin{eqnarray}
t_{\rm md} & = & \frac{3 c^3 I}{B_p^2 R^6 \Omega_i^2} \simeq (2.0 \times 10^3~{\rm s}) I_{45} B_{p,15}^{-2} P_{i,-3}^2 R_6^{-6}, \\
t_{\rm GW} & = & \frac{5c^5}{128 GI \epsilon^2 \Omega_i^4} \simeq (9.1\times 10^3~{\rm s}) I_{45}^{-1} P_{i,-3}^4 \epsilon_{-3}^{-2},
\end{eqnarray}
where $\Omega_i$ and $P_i$ are the initial angular velocity and period of the magnetar, respectively. Hereafter the convention $Q_n=Q / 10^n$ has been adopted.
Before the spin-down time, $\Omega$ and therefore $L_{\rm sd}$ is essentially a constant. With dipolar spin-down and GW spin-down terms only, the decay of $\Omega$ after their respective spin-down time scale goes as $\Omega \propto t^{-1/2}$ and $\propto t^{-1/4}$, respectively, so that $L_{\rm sd} \propto t^{-2}$ and $\propto t^{-1}$ respectively. When both dipolar and GW spin-down terms are considered, the overall spin-down time scale of the magnetar
can be defined as
\begin{equation}
t_{\rm sd} = {\rm min} (t_{\rm md}, t_{\rm GW}),
\label{eq:t_sd}
\end{equation}
and the spin-down luminosity (Eq.(\ref{eqs:lsd})) behaves as
\begin{equation}
 L_{\rm sd}  \propto  \left\{
  \begin{array}{ll}
     t^0, & t<t_{\rm sd}=t_{\rm GW} , \\
     t^{-1}, & t_{\rm sd}=t_{\rm GW} < t < t_{\rm md}, \\
     t^{-2}, & t > t_{\rm md}
  \end{array}
  \right.
\end{equation}
for $t_{\rm GW} < t_{\rm md}$, and 
\begin{equation}
 L_{\rm sd}  \propto  \left\{
  \begin{array}{ll}
     t^0, & t<t_{\rm sd}=t_{\rm md}, \\
     t^{-2}, & t >t_{\rm sd}= t_{\rm md}
  \end{array}
  \right.
\end{equation}
for $t_{\rm md} < t_{\rm GW}$.

In the free zone, internal dissipation of the magnetar wind gives an X-ray luminosity which scales with $L_{\rm sd}$, i.e.
\begin{equation}
L_{\rm X,free}(t)=\eta L_{\rm sd} = \frac{\eta B_p^2 R^6 \Omega^4(t)}{6c^3},
\label{eqs:X-free}
\end{equation}
where $\eta$ is the efficiency of converting the dipole spin down luminosity to the observed X-ray luminosity.

\subsection{Trapped zone}

In the trapped zone, the dissipated photon energy defined by Eq.(\ref{eqs:lsd}) is trapped in the ejecta. Together with the non-dissipated Poynting flux energy, it is used to heat the ejecta and accelerate the ejecta via $p dV$ work. The optical depth of the ejecta is written as 
\begin{equation}
\tau=\kappa (M_{\rm ej}/V')(R/\Gamma),
\end{equation} 
where $\kappa$ is the opacity of the ejecta, and $V'$ is the co-moving volume.
The ejecta is initially opaque when $\tau > 1$, so that X-ray emission is essentially the Wien tail of the merger-nova photosphere emission, if there is no significant energy dissipation and Comptonization below the photosphere (which we assume for simplicity in this paper). The trapped magnetar wind however becomes transparent after $t_\tau$ defined by $\tau = 1$. The X-ray luminosity would rise up quickly to the level of free zone luminosity at $t > t_\tau$ (Eq.(\ref{eqs:X-free})) if the magnetar is still alive, since the newly dissipated X-rays would escape the remnant without being reprocessed.

In order to calculate the X-ray lightcurve, one needs to solve the dynamical evolution of the merger-nova ejecta, which is heated by radioactive decay and energy injection from the magnetar \citep{yu13}.

The total energy of the ejecta excluding the rest mass energy can be expressed as
\begin{equation}
E_{\rm ej}=(\Gamma-1)M_{\rm ej}c^2+\Gamma E'_{\rm int}
\label{eqs:E_ej}
\end{equation}
where $\Gamma$ is the Lorentz factor, $E'_{\rm int}$ is the internal energy in the co-moving frame, and the two terms represent the kinetic energy and the thermal energy of the ejecta, respectively. For each time step $dt$, the ejecta receives energy from the magnetar (with luminosity $L_{\rm sd}$) and from radioactive decay of the neutron-rich materials (with luminosity $L_{\rm ra}$, and in the mean time lose energy through emission of the electrons (with luminosity $L_e$).  
Energy conservation therefore gives
\begin{equation}
dE_{\rm ej}=(L_{\rm sd}+L_{\rm ra}-L_{e})dt,
\label{eqs:dE_ej}
\end{equation}
where $t$ is the observer's time, $ L_{\rm sd} $ is defined by Eq.(\ref{eqs:lsd}), $ L_{\rm ra} $ is the radioactive power, and $ L_{e} $ is bolometric radiation luminosity by the heated electrons. Equating the derivative of Eq.(\ref{eqs:E_ej}) with Eq.(\ref{eqs:dE_ej}) and noticing that $d t' = {\cal D} dt$ ($t'$ is the co-moving time), where $ {\cal D}=1/[\Gamma(1-\beta\cos\theta)] $ is the Doppler factor with $ \beta=\sqrt{1-\Gamma^{-2}} $ and $\theta=0$ for an on-beam observer), one has
\begin{equation}
\frac{d\Gamma}{dt}=\frac{L_{\rm sd}+L_{\rm ra}-L_{e}-\Gamma{\cal D}(dE'_{\rm int}/dt')}{M_{\rm ej}c^2+E'_{\rm int}}.
\end{equation}
The change of the internal energy includes heating from magnetar and radioactivity and cooling due to radiation and $pdV$ work \citep{KB10}. Therefore, in the co-moving frame, the evolution of internal energy can be calculated as \citep{yu13}
\begin{equation}
\frac{dE'_{\rm int}}{dt'}=\xi L'_{\rm sd}+L'_{\rm ra}-L'_e-P'\frac{dV'}{dt'}
\end{equation}
where $\xi$ is an efficiency parameter to define the fraction of the spin-down energy that is used to heat the ejecta. The co-moving luminosities are defined as $ L'_{\rm sd}=L_{\rm sd}/{\cal D}^2 $, $ L'_{\rm ra}=L_{\rm ra}/{\cal D}^2 $ 
and $ L'_{e}=L_{e}/{\cal D}^2 $, and the co-moving radiative heating luminosity can be calculated as 
\begin{eqnarray}
\lefteqn{L'_{\rm ra}=4\times 10^{49}M_{\rm ej,-2}}
\nonumber\\
& & \times\left[\frac{1}{2}-\frac{1}{\pi}\arctan\left(\frac{t'-t'_0}{t'_{\sigma}} \right) \right]^{1.3}{\rm erg~s^{-1}}.
\end{eqnarray}
with $t'_{0}\sim1.3$ $\rm s$ and $t'_{\sigma}\sim 0.11$ $\rm s$ \citep{korobkin12}. For a relativistic gas, the pressure is (1/3) of the internal energy density, i.e.
\begin{equation}
P'=E'_{\rm int}/(3V'). 
\end{equation}
The co-moving volume is determined by
\begin{equation}
\frac{dV'}{dt'}=4\pi R^2\beta c,
\label{eq:vol}
\end{equation}
and the effective velocity of the ejecta for the observer can be generally written as (valid for both non-relativistic and relativistic regimes)
\begin{equation}
\frac{dR}{dt}=\frac{\beta c}{1-\beta}.
\end{equation}

The co-moving frame bolometric emission luminosity of the heated electrons can be estimated as
\begin{equation}
L'_e=\left \{ \begin{array}{ll} 
E'_{\rm int}c/(\tau R/\Gamma), & \textrm{for $t < t_{\tau}$},\\
\\
E'_{\rm int}c/(R/\Gamma), & \textrm{for $t \geq t_{\tau}$},\\
\end{array} \right.
\end{equation}
where the first expression takes into account the skin-depth effect of an optically thick emitter.
The observed spectrum is nearly blackbody with a typical temperature
\begin{equation}
\varepsilon_{\gamma,p}\approx 4{\cal D}kT'=\left \{ \begin{array}{ll} 
4{\cal D}k\left( \frac{E'_{\rm int}}{aV'\tau}\right) ^{1/4} & \textrm{for $\tau>1$}\\
\\
4{\cal D}k\left( \frac{E'_{\rm int}}{aV'}\right) ^{1/4}  & \textrm{for $\tau \leq 1$}\\
\end{array} \right.
\end{equation}
where $k$ is the Boltzmann constant and $a$ is the blackbody radiation constant. 

For a blackbody spectrum with co-moving temperature $T'$, the luminosity at a particular frequency $\nu$ 
is given by
\begin{equation}
(\nu L_{\nu})_{\rm bb}=\frac{8\pi^2 {\cal D}^2R^2}{h^3c^2}\frac{(h\nu /{\cal D})^4}{exp(h\nu/{\cal D}kT')-1}.
\label{eq:Lmn}
\end{equation}

So for a stable magnetar, the observed X-ray luminosity in the trapped zone can be written as
\begin{equation}
L_{\rm X,trapped}(t)= e^{-\tau} \frac{\eta B_p^2 R^6 \Omega^4(t)}{6c^3} + (\nu_{\rm X} L_{\rm \nu,X})_{\rm bb},
\label{eqs:X-free}
\end{equation}
where the first term is emission from the dissipating wind, which is negligible when $\tau \gg 1$, and
the second term is the Wien tail of the merger-nova photosphere, which can be calculated using
Eq.(\ref{eq:Lmn}) by choosing a typical X-ray frequency.

\subsection{Collapse of the supra-massive neutron star}

The final product of a NS-NS merger depends on the initial masses of the two pre-merger NSs and the EoS of the compact object. For a merger product with mass $M_s$ and a given EoS, one may define a critical period $P_c$ above which the NS would collapse into a black hole. For a given EoS, the maximum mass $M_{\rm max}$ a NS can sustain is a function of period $P$. The shorter the $P$, the larger the $M_{\rm max}$, which can be cast into the form \citep{lasky14,RL14,li16c}
\begin{equation}
M_{\rm max} = M_{\rm TOV} (1+\alpha P^{\beta}),
\label{eq:M_max}
\end{equation}
where $M_{\rm TOV}$ is the maximum mass of a NS with zero spin for a given EoS, $\alpha$ and $\beta$ are the phenomenological parameters to describe how the maximum mass of a supra-massive NS depends on $M_{\rm TOV}$ and $P$, which depends on the NS EoS. Notice that $\beta < 0$ guarantees a decreasing $M_{\rm max}$ with increasing $P$. The collapse time can be calculated by equating $M_s$ and $M_{\rm max}$, which gives 
\begin{equation}
P_c = \left(\frac{M_s-M_{\rm TOV}}{\alpha M_{\rm TOV}}\right)^{1/\beta}.
\end{equation}
The merger product is a supra-massive NS if $P_c > P_i$, but would be a prompt BH (direct collapse or hypermassive NS) if $P_c < P_i$.
Using Eq.(\ref{eq:Edot}), one can derive the collapse time \citep{gao16}
\begin{equation}
t_{\rm col} = \frac{a}{2 b^2} \ln \left[ \left( \frac{a \Omega_i^2 + b}{a \Omega_c^2+b}\right) \frac{\Omega_c^2}{\Omega_i^2} \right] + \frac{\Omega_i^2-\Omega_c^2}{2b\Omega_i^2 \Omega_c^2},
\label{eq:t_col}
\end{equation}
where $\Omega_c = 2\pi/P_c$, $a= (32 GI\epsilon^2) / (5c^5)$ and $b=(B_p^2 R^6)/(6c^3I)$, and $a$ and $b$ are assumed to be approximately constant during the evolution. 

The evidence of direct collapse of a supra-massive NS to a BH is collected from the sGRB observations. By analyzing the joint BAT-XRT light curves of sGRBs, \cite{lv15} found that a good fraction of them show an internal plateau (the temporal segment with decay slope close to flat), followed by a decay slope steeper than 3. This implies internal dissipation of a magnetar central engine with the steep decay marking the collapse of the supra-massive NS into a black hole. \cite{gao16} obtained a minimum $22\%$ fraction of supra-massive NSs as the central engine of sGRBs. Within the EoS GM1 \citep{lasky14}, \cite{gao16} obtained $\sim$ 30$\%$ supra-massive NSs that collapses to black holes in a range of delay time scales, $\sim 30\%$ stable NSs, and $\sim 40 \%$ prompt black holes if the cosmological NS-NS systems follow the mass distribution of Galactic NS-NS systems \citep{kiziltan13,martinez15}. 

\subsection{Equations of State}

The outcome of our model depends on the EoS of the central compact star, which is not uniquely constrained either theoretically or observationally. 
Recently, \cite{li16c} studied the spin-dependent compact star structure for four recently-constructed ``unified'' NS EoSs (BCPM, BSk20, BSk21, and Shen) and three developed strange quark star (QS) EoSs (CIDDM, CDDM1, and CDDM2) and derived the $\alpha$ and $\beta$ values of each EoS in the convention of Eq.(\ref{eq:M_max}). They found that except BCPM, all other six EoSs can reproduce the observed supra-massive NS/QS fraction. Furthermore, they found that the QS EoSs can better reproduce the observed collapse time distribution as inferred from the data \citep{lv15}. The best match with the data constrains the distributions of different parameters ($P_i$, $B_p$, $\epsilon$, and $\eta$) for different EoSs (Table I.II of \citep{li16c}). Again assuming the Galactic NS-NS mass distribution, we calculate the post-merger product fractions, which are shown in Table \ref{tab:eos}. These results are used below to construct the peak luminosity function of these X-ray transients.

\renewcommand\arraystretch{1.5}
\begin{table}[hbtp]
\caption{Post-merger product fractions for different EoSs.}
\centering
\begin{tabular}{c|cccc|c|c}
\hline 
\hline
 & $ M_{\rm TOV}$ & $\alpha$ & $\beta$ & Ref.& $f_{\rm BH}:f_{\rm SMNS}:f_{\rm SNS}$ & Type \\
 & $(M_{\odot})$ & $(P^{-\beta})$ && & & \\ 
\hline 
GM1 & 2.37 & 0.0523 & -2.840 & (1) &$40\%:30\%:30\% $ & NS \\ 
\hline 
BSk20 & 2.17  & 0.0359 & -2.675 & (2) & $72\%:26\%:2\% $ & NS \\ 
\hline 
BSk21 & 2.28 & 0.0487 & -2.746 & (2)& $20\%:70\%:10\% $ & NS \\ 
\hline 
Shen & 2.18 & 0.0766 & -2.738  & (2)&$10\%:87\%:3\% $ & NS \\ 
\hline 
CIDDM & 2.09 & 0.1615 & -4.932  & (2)& $30\%:70\%:- $ & QS \\ 
\hline 
CDDM1 &  2.21  & 0.3915 & -4.999 & (2)& $ -:96\%:4\% $ & QS \\ 
\hline 
CDDM2 &  2.45  & 0.7448 & -5.175  & (2)& $ -:50\%:50\%$ & QS \\ 
\hline
\hline
\end{tabular}
\tablerefs{(1).\cite{lasky14}; (2).\cite{li16c}.}
\label{tab:eos} 
\end{table}

\subsection{Characteristic times \& gallery of light curves}

The lightcurves of the X-ray transients depend on the relative positions among the following three critical times: 
\begin{itemize}
 \item $t_{\rm sd}$: the spin-down time of NSs, as defined in Eq.(\ref{eq:t_sd});
 \item $t_{\tau}$: the time at which $\tau=1$ is satisfied, which is the transition point of wind emission from being trapped to free;
 \item $t_{\rm col}$: the time at which the NS collapses into black hole (Eq.(\ref{eq:t_col})).
\end{itemize}

\begin{figure*}[]
\centering
\includegraphics[width=0.3\textwidth]{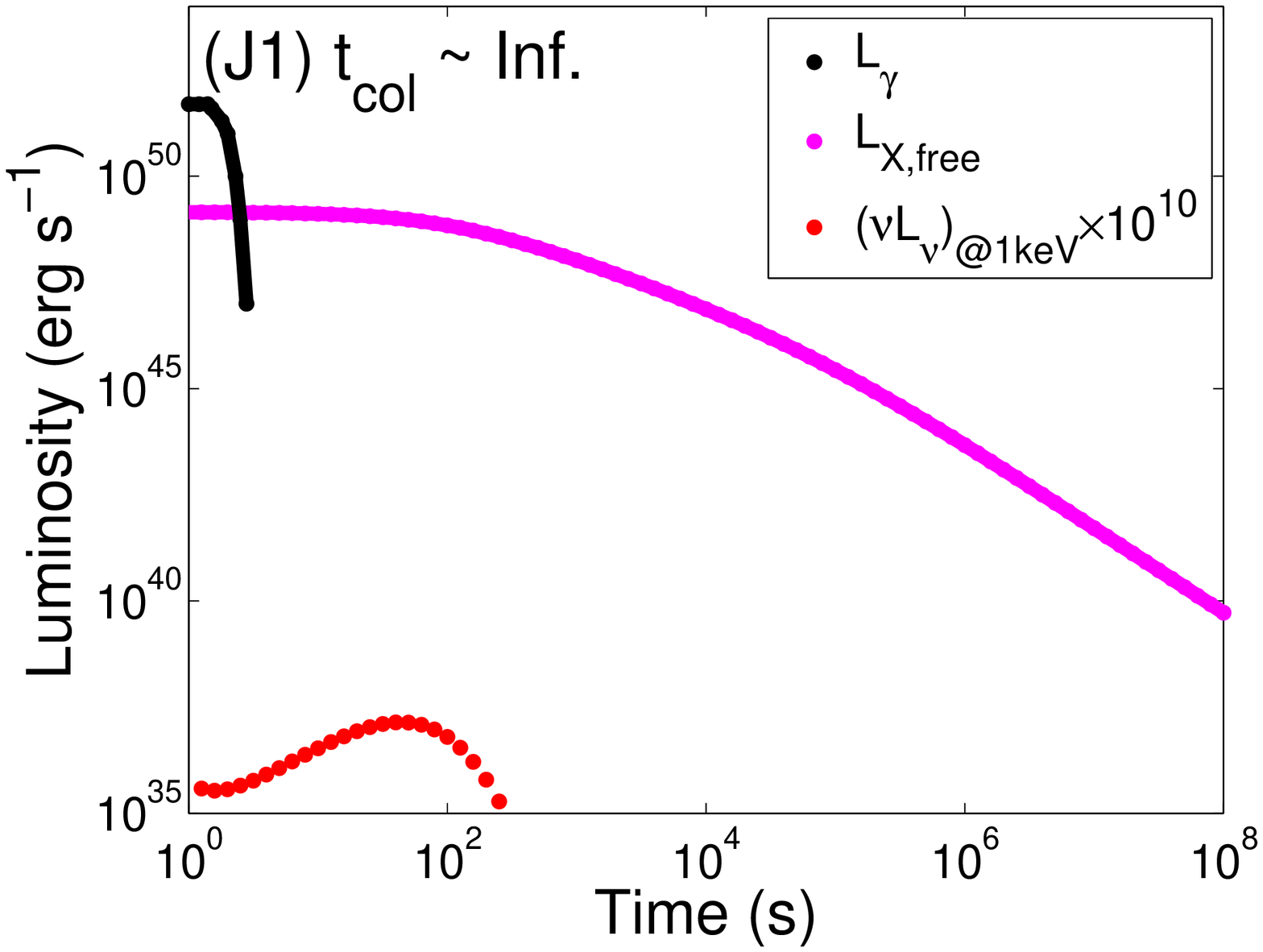}
\includegraphics[width=0.3\textwidth]{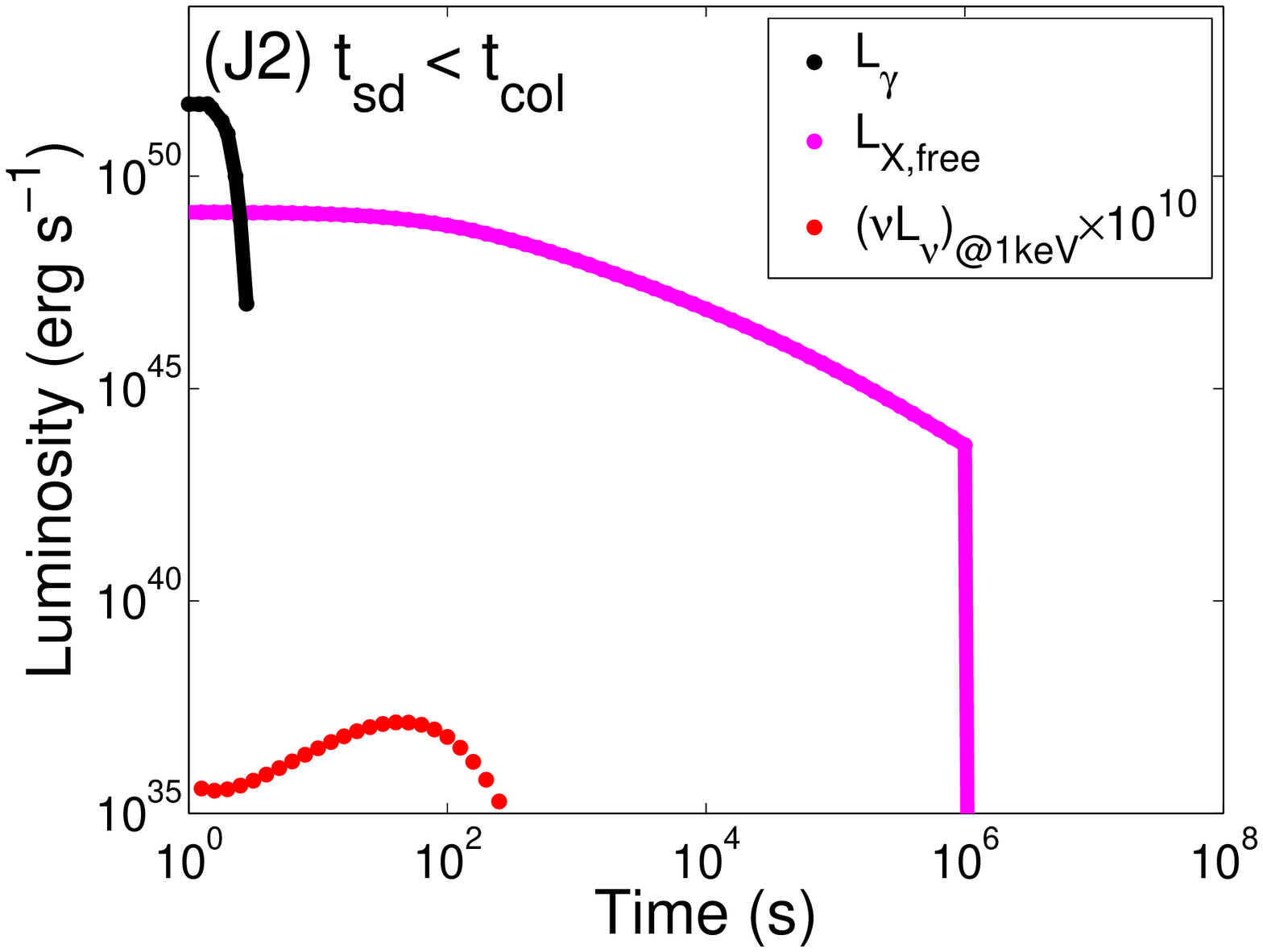}
\includegraphics[width=0.3\textwidth]{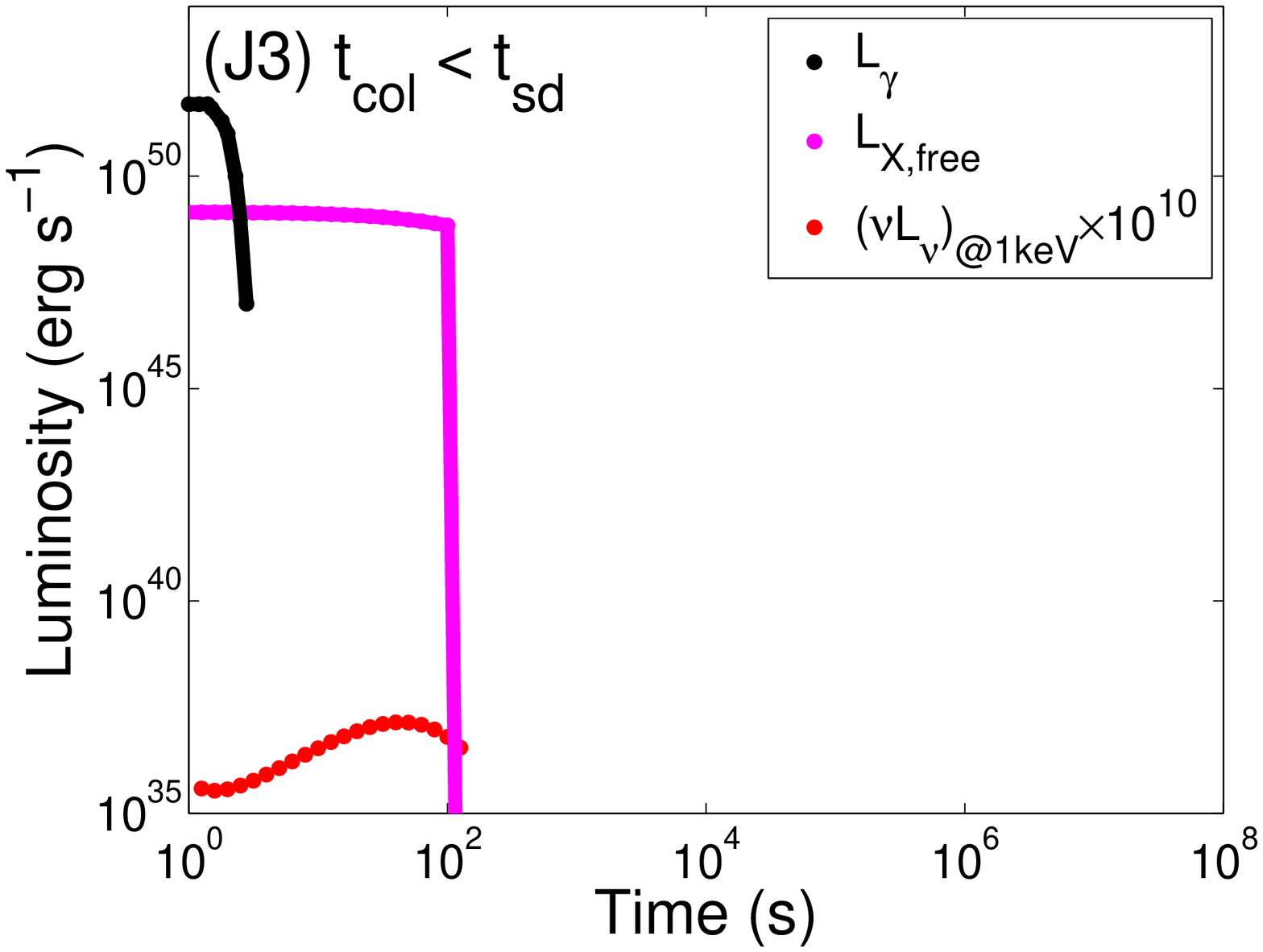}
\includegraphics[width=0.3\textwidth]{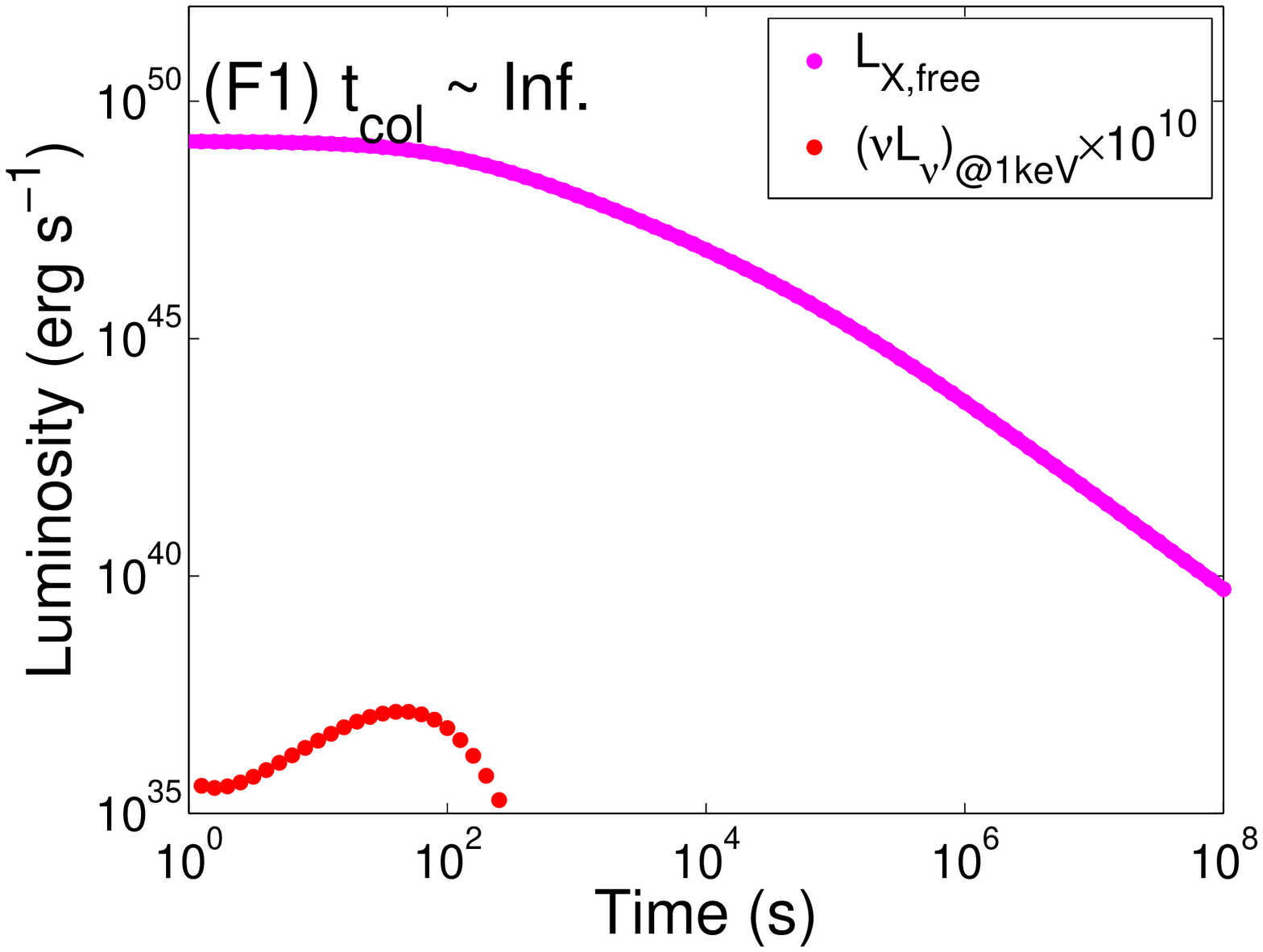}
\includegraphics[width=0.3\textwidth]{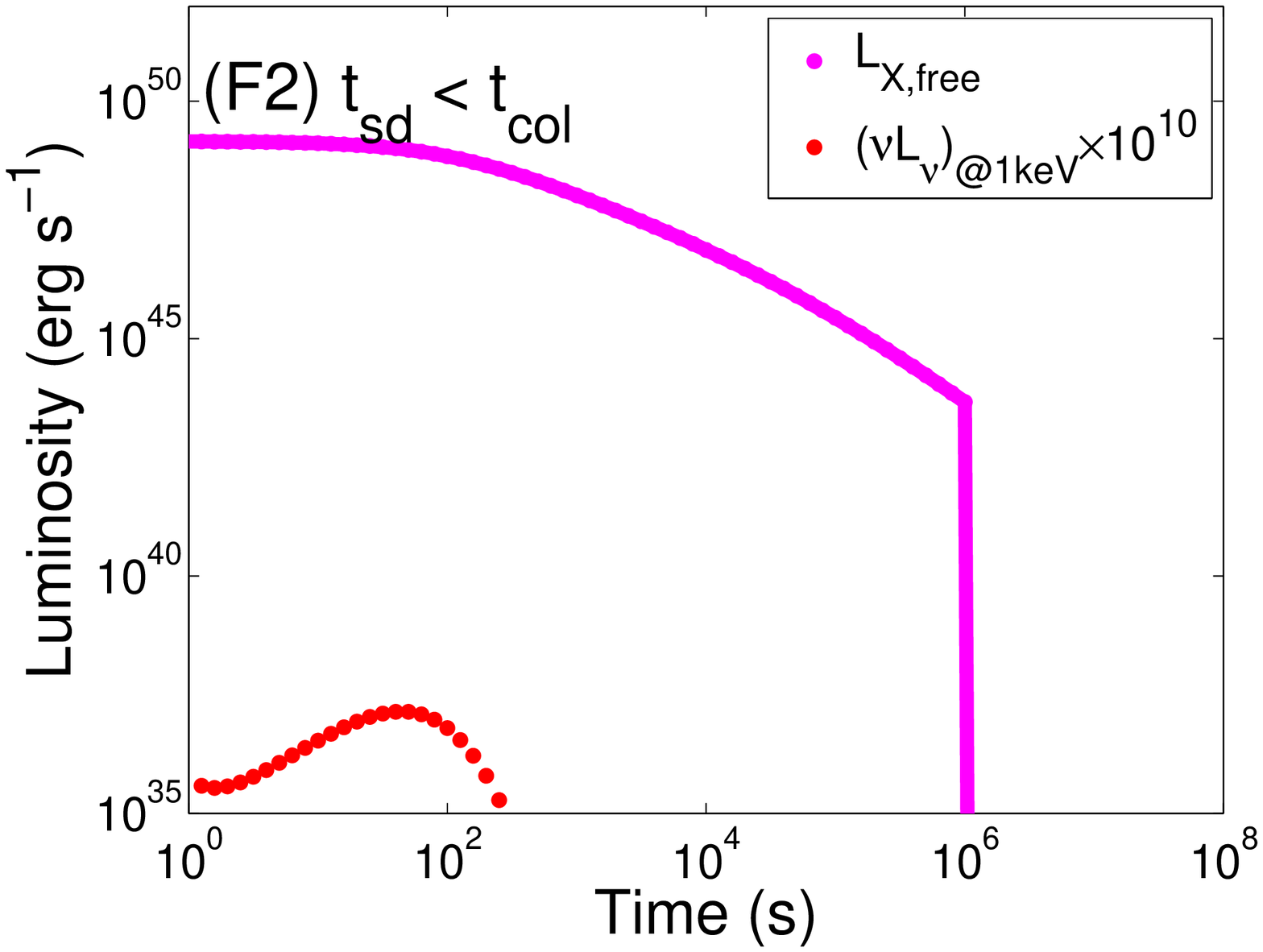}
\includegraphics[width=0.3\textwidth]{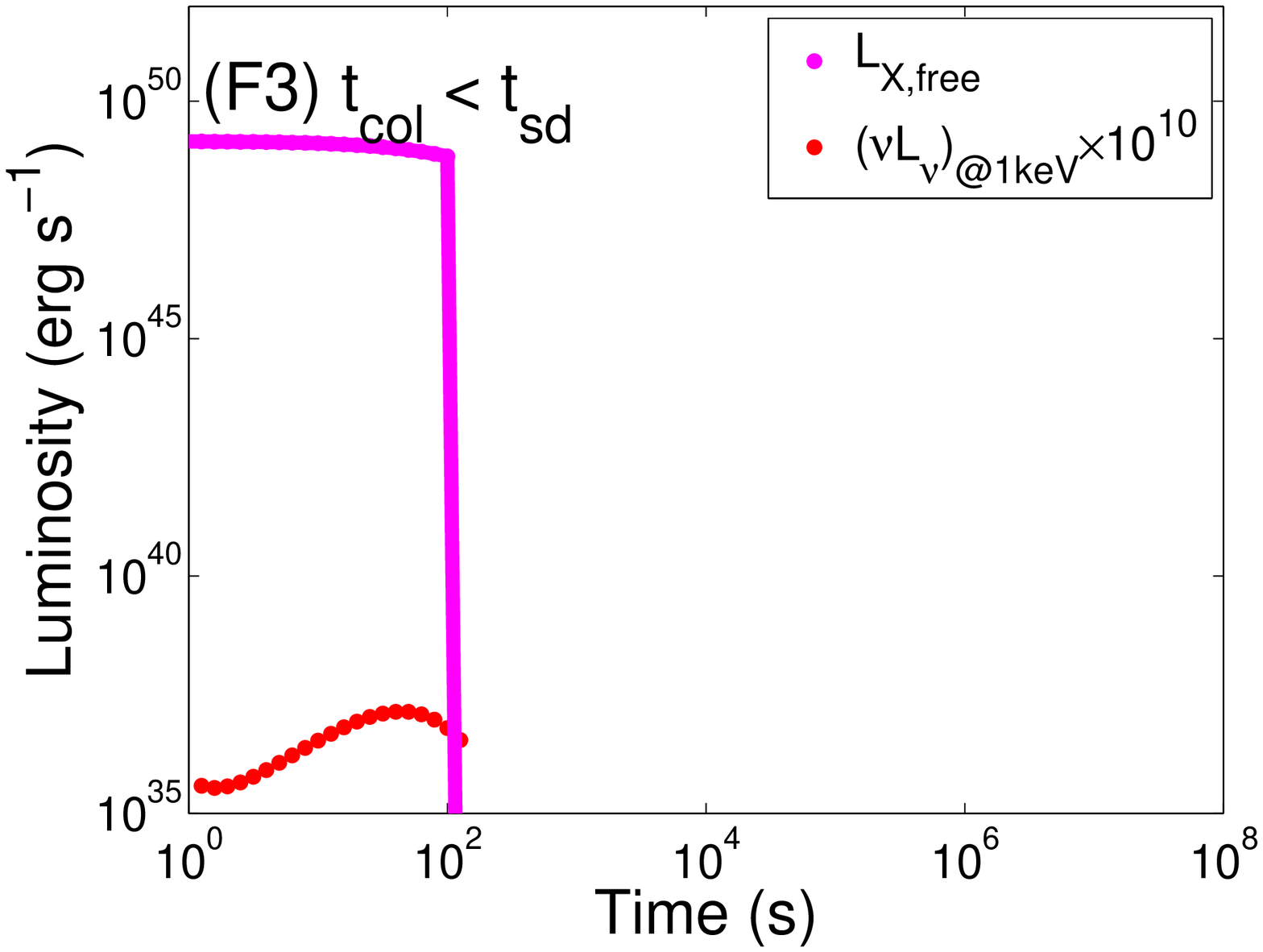}
\includegraphics[width=0.3\textwidth]{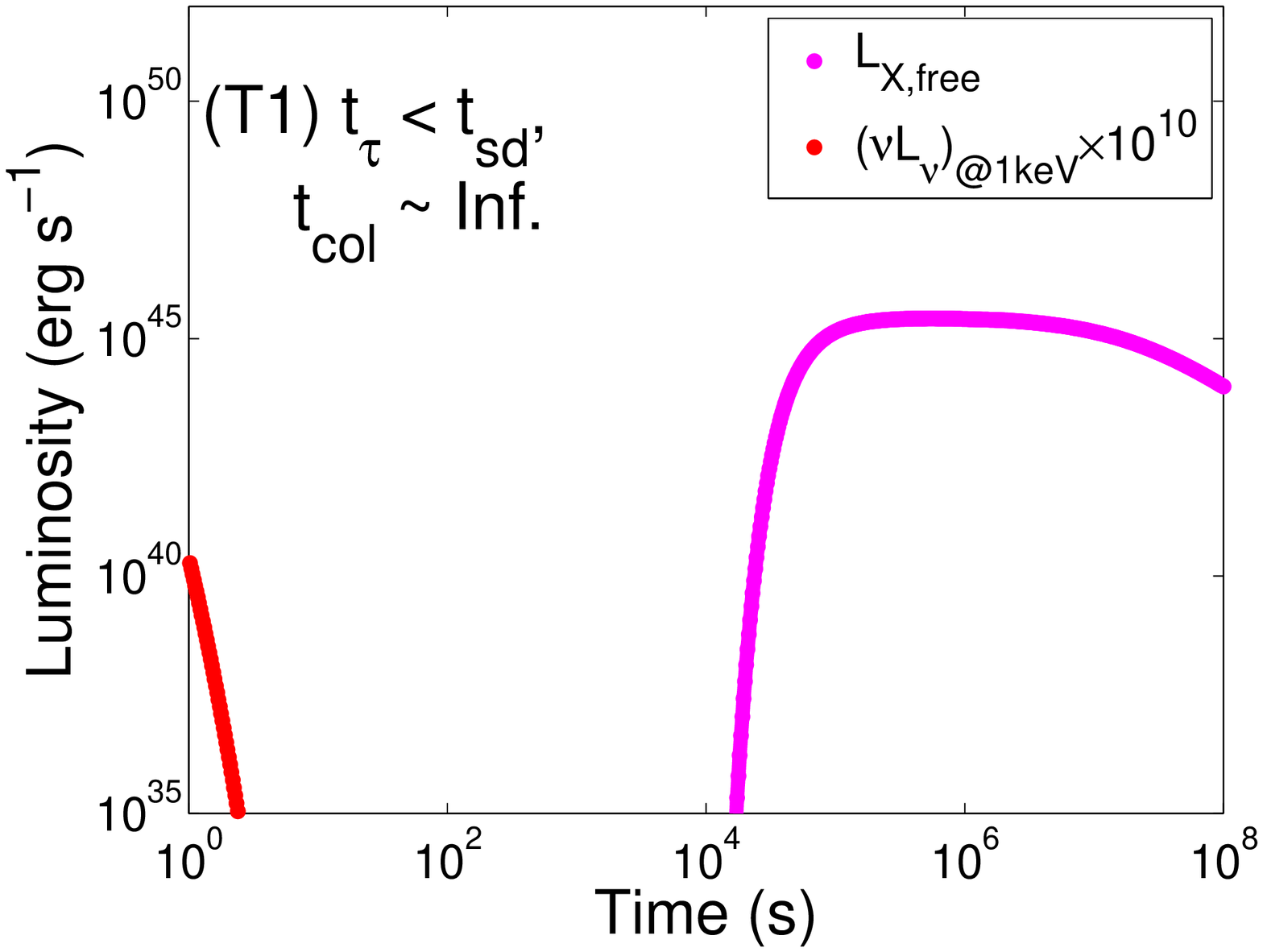}
\includegraphics[width=0.3\textwidth]{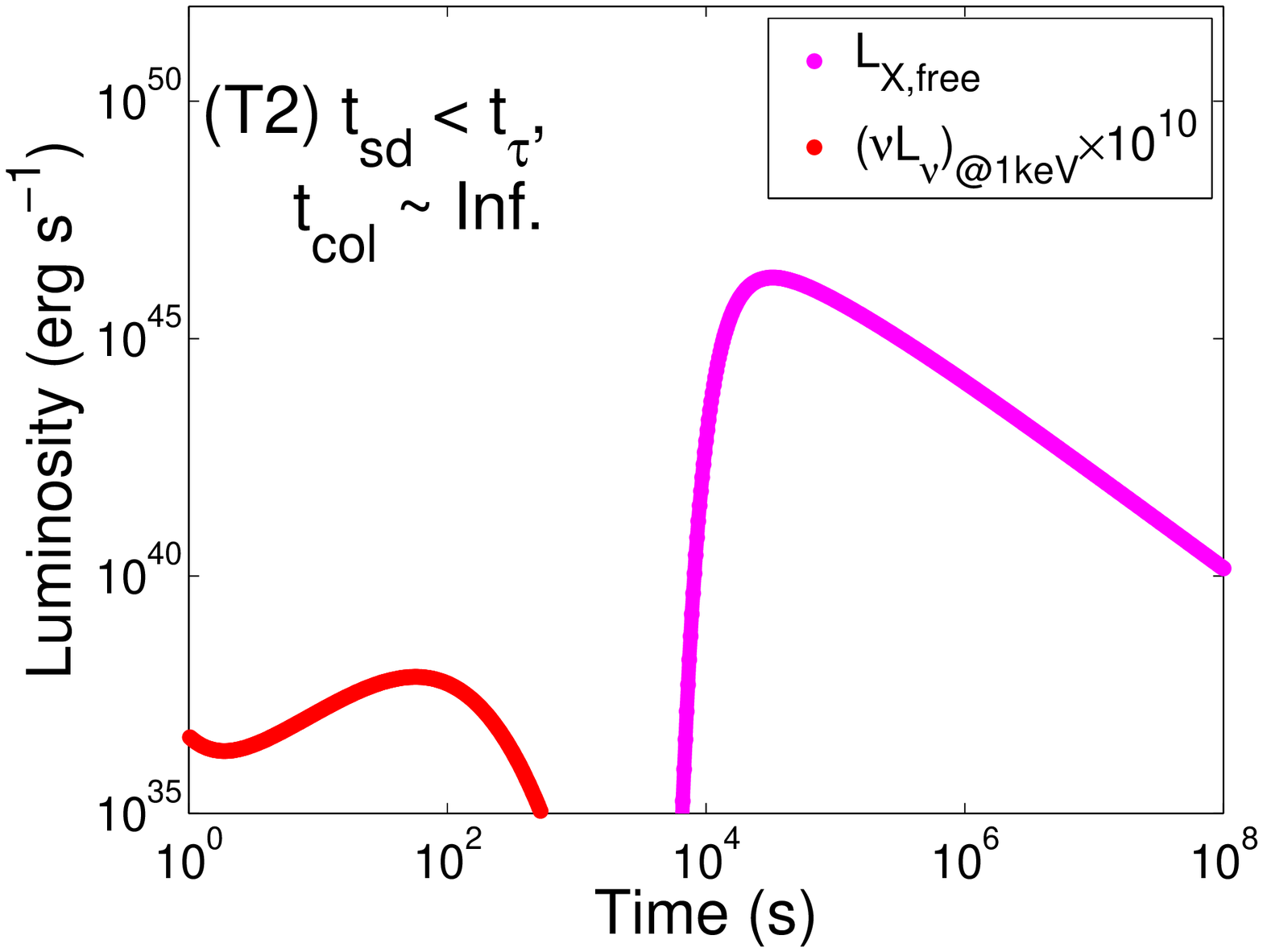}
\includegraphics[width=0.3\textwidth]{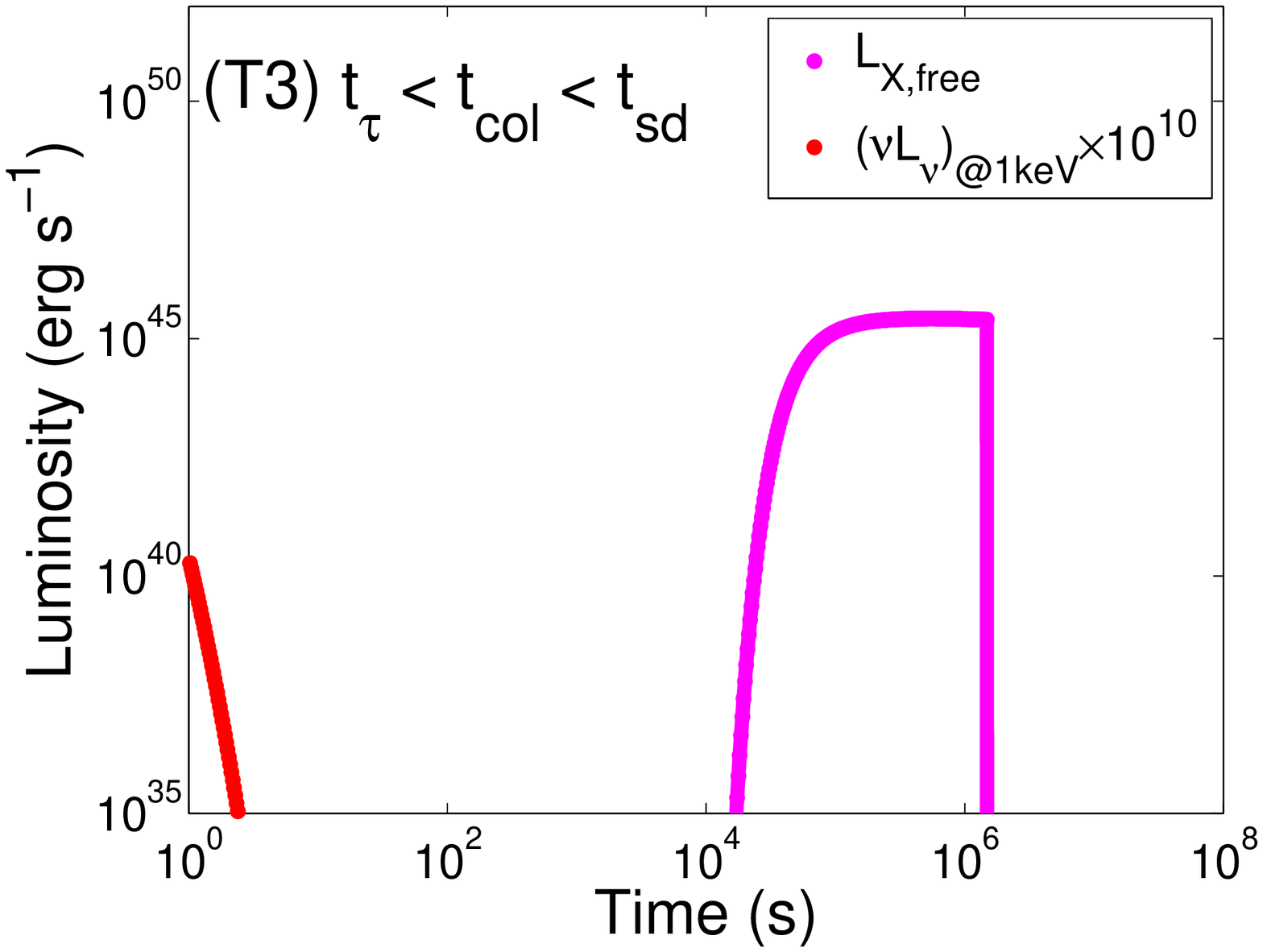}
\includegraphics[width=0.3\textwidth]{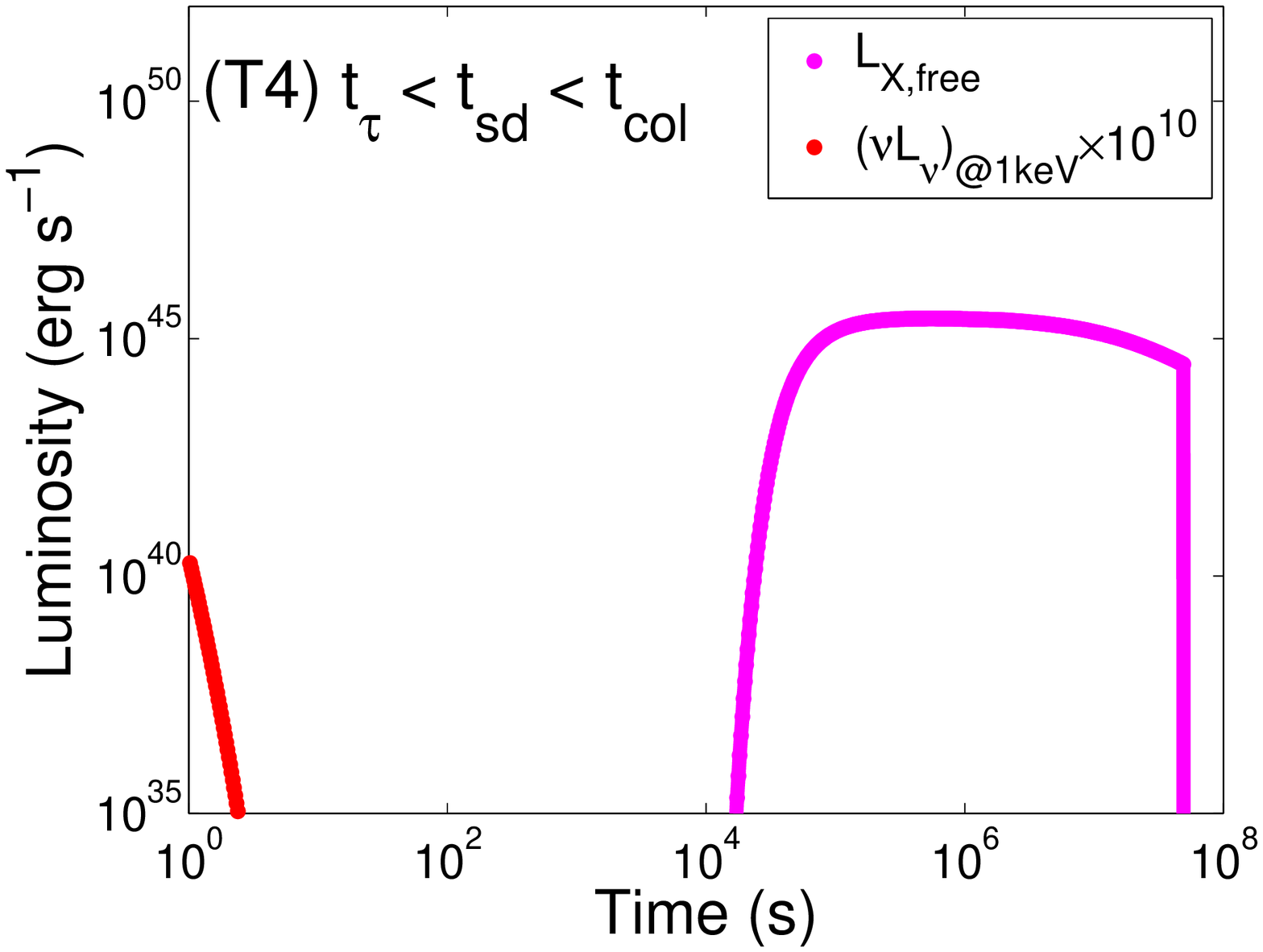}
\includegraphics[width=0.3\textwidth]{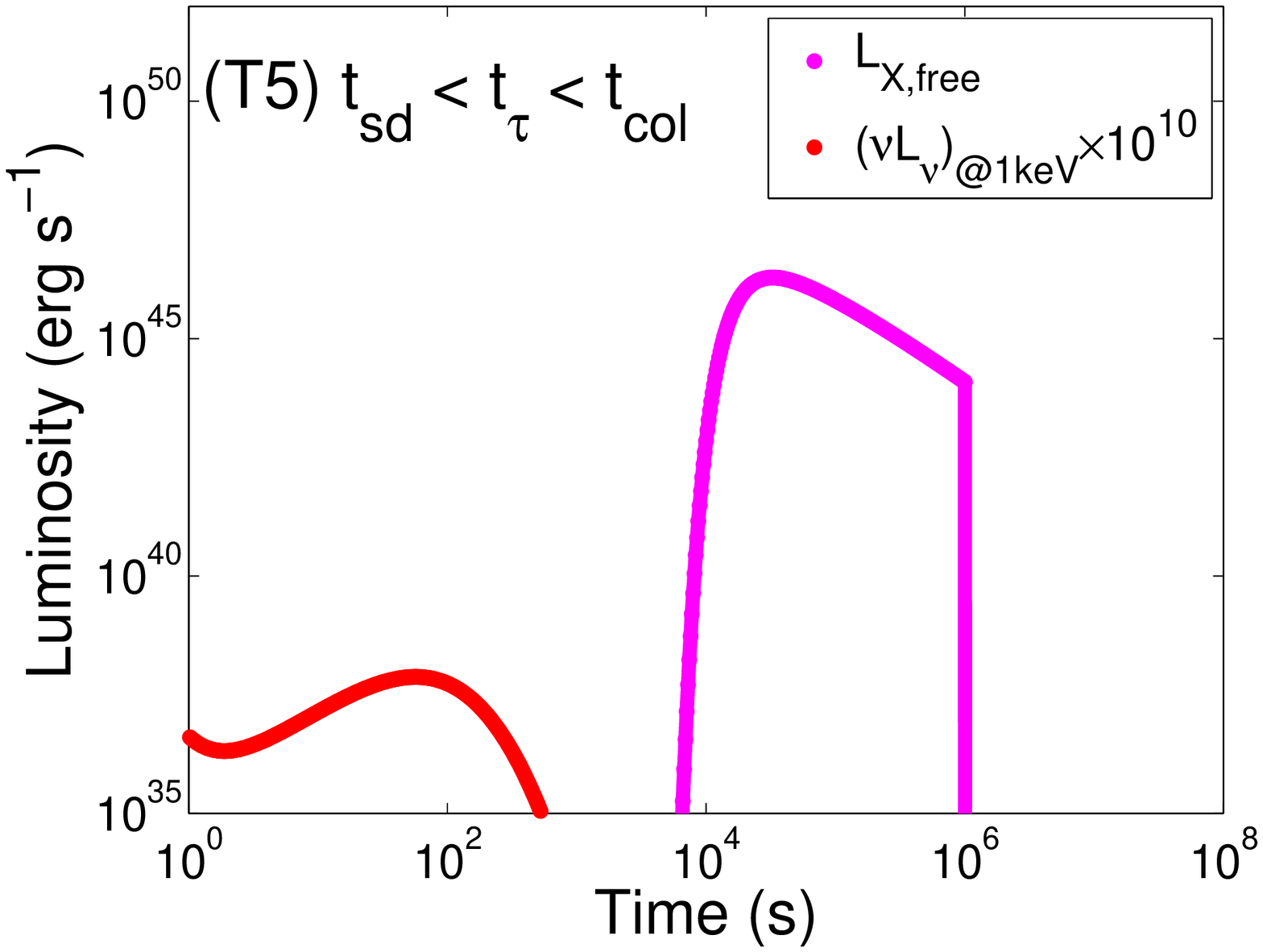}
\includegraphics[width=0.3\textwidth]{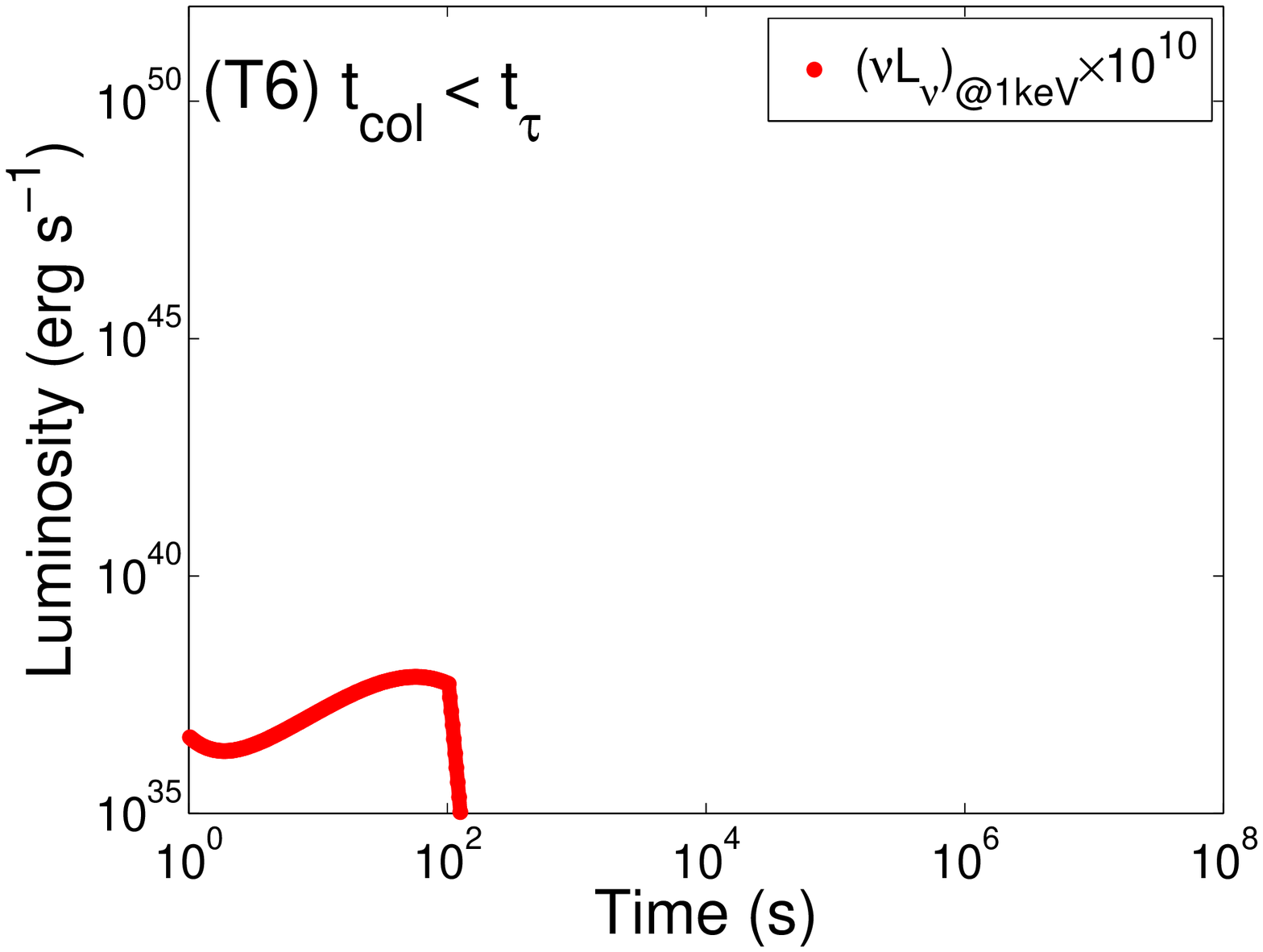}
\caption{A gallery of all possible X-ray light curves. Both the wind emission (magenta) and X-ray merger-nova (red, boosted by $10^{10}$) are presented, with solid lines showing the observed flux (given unlimited sensitivity) and dashed lines showing the merger-nova emission outshone by the wind emission.  (J1-3): jet zone light curves (the sGRBs is also plotted (black)); (F1-3): free zone light curves; (T1-6): trapped zone light curves.}
\label{Fig:lc}
\end{figure*}

A gallery of all 12 possible X-ray lightcurves is collected in Figure \ref{Fig:lc}. There are three general types that correspond to the three observational zones: jet zone (J1-J3), free zone (F1-F3) and trapped zone (T1-T6).

The jet zone and free zone lightcurves are similar to each other, except that there is a short GRB observed (marked as the black solid line) for the jet zone but not in the free zone. There are three possible lightcurves for each case, which correspond to the cases of a stable NS (J1 and F1), a supra-massive NS with collapse happening during the decay phase ($t_{\rm col} > t_{\rm sd}$, J2 and F2) and during the plateau phase ($t_{\rm col} < t_{\rm sd}$, J3 and F3), respectively. Given the typical magnetar parameters (see section 3.1 for details), the plateau is as bright as $10^{49-50}$ $\rm erg\,s^{-1}$, which is also consistent with the luminosity of the internal X-ray plateaus seen in some sGRB afterglows. The merger-nova typically peaks in optical/IR. In X-rays (dotted line) it is much fainter than the direct wind dissipation component, i.e. below $10^{40}$ $\rm erg\,s^{-1}$, so that it is not detectable. For indicative purposes, we mark a cutoff of lightcurve at $t_{\rm col}$. In reality, the decay should be shallower due to the so-called ``curvature effect'', i.e. the delay of arrival of high-latitude photons \citep{kumar00,uhm15}, but due to the effect of the zero time offset \citep{zhang06} and a possible ``bulk acceleration" effect \citep{uhm15,uhm16}, the actual decay slope can be very steep, e.g. $\propto t^{-10}$ as seen in the observations \citep{rowlinson10,rowlinson13,lv15}. In the jet zone, interaction between the GRB jet and the ambient medium would also power a bright X-ray afterglow, which may appear as an external plateau \citep[e.g.][]{lv15}. However, such afterglow emission is diminished in the free zone. The dominant X-ray emission should come from the nearly isotropic internal dissipation emission of the magnetar.

In the trapped zone, the lightcurves are more complicated, as shown in T1-6 in Figure.\ref{Fig:lc}. It includes two components: the merger-nova component (mostly too faint to be detected) and the wind dissipation component. Due to the $e^{-\tau}$ factor, the wind emission can diffuse out only when the ejecta gets transparent around $t_\tau$ \citep[e.g.][]{gao15}.  If the collapse happens before the ejecta becomes optically thin, there is no wind emission observed (T6). Otherwise, the wind emission would come out at around $t_\tau$.
In the case of a stable NS ($t_{\rm col} =$ infinity), the wind emission always comes out, either during the plateau phase ($t_\tau < t_{\rm sd}$, T1) or during the decay phase ($t_\tau > t_{\rm sd}$, T2). In the case of supra-massive NSs, one has four cases depending on the comparison among $t_{\rm col}$, $t_\tau$ and $t_{\rm sd}$: for $t_\tau < t_{\rm col} < t_{\rm sd}$ (T3), one observes part of the plateau and the star collapses before spinning down; for $t_\tau < t_{\rm sd} < t_{\rm col}$ (T4), one observes part of the plateau and a decay segment before the collapse; for $t_{\rm sd} < t_\tau < t_{\rm col}$ (T5), one can only observe a decay segment before the collapse; and for $t_{\rm col} < t_\tau$ (T6), the wind emission cannot be observed. For typical parameters, the merger-nova component (red solid line) for all 6 cases is too faint to be detected (as shown in Fig. \ref{Fig:lc}), but with extreme parameters (e.g. $B_p > 10^{16}$ G), the X-ray emission of the merger-nova can be very bright \citep{siegel16a,siegel16b}.

To sum up, bright X-ray transients are expected from the free zone, while in the trapped zone the chance of observing bright events is much lower. As a result, based on the current non-detection of bright X-ray transients one may constrain the solid angle ratio parameter $k_\Omega$ (Eq.(\ref{eq:k_Omega})) for the free zone.
One can also predict the luminosity function and event rate density of these transients once $k_\Omega$ and EoS are determined and study the detectability of these transients by current and future wide-field X-ray or soft $\gamma$-ray detectors.

\section{Monte Carlo simulations: peak luminosity function \& event rate density}

\subsection{Simulations}

The luminosity function and event rate density of the X-ray transients depend on many factors, including the EoS, solid angle fraction $k_\Omega$, and many other unknown parameters associated with the ejecta and the magnetar. Lacking observational data, it is impossible to give a unique prediction. In the following, we take GM1 EoS as an example \citep{lasky14,gao16}, and discuss other EoSs through a comparison of the results. For each EoS, we adopt some typical parameters inferred from the sGRB data, and make predictions by assuming different values of $k_\Omega$. 

For the GM1 EoS \citep{GM91} , we adopt $M_{\rm TOV}=2.37M_{\odot}$, $R=12.05\rm km$, $I=2.13\times10^{45}\rm g\,cm^{-2}$, $\alpha=1.58\times10^{-10}s^{-\beta}$ and $\beta=-2.84$ using the prescription of Eq.(\ref{eq:M_max}) \citep{lasky14}. The following parameters are adopted in the simulations based on previous work \citep{gao16}: the ejecta mass with a lognormal distribution $N_{\rm ej}$($\mu_{\rm ej}=10^{-2}M_{\odot},\sigma_{\rm ej}=0.5)$, the dipolar magnetic field strength with a lognormal distribution $N_B$($\mu_{B}=10^{15}G,\sigma_{B}=0.2$), the initial period $P_i = 1$ ms, ellipticity of the nascent NS $\epsilon=0.005$, and efficiency parameters $\xi=0.5$, $\eta=0.5$. For other EoSs, the best-fit parameters derived from \cite{li16c} are adopted. For all the simulations, the opacity parameter is adopted as $\kappa = 2\, {\rm cm^{2}\,g^{-1}}$. This is an unknown parameter, which ranges from 0.1-10 ${\rm cm^{2}\,g^{-1}}$ depending on whether lanthanides dominate the opacity \citep{BK13,TH13,MP14}. Our moderate value is based on the consideration that the lanthanides do exist in the ejecta, on the other hand, the magnetar wind tends to destroy the heavy elements and reduce opacity. In reality, different merger systems may have different $\kappa$ values, and our typical value may be regarded as the average value of opacity among different events. We simulate 10000 events for the trapped-zone events. The number of free-zone events can be correspondingly simulated based on the assumed $k_\Omega$.

\subsection{Peak luminosity function \& event rate density: \\the case of GM1}

The peak luminosity of the X-ray transients depends on several factors. In the free zone, it is the luminosity during the plateau phase. In the trapped zone, if $t_{\rm col} > t_\tau$, the peak luminosity is simply Eq.(\ref{eqs:X-free}) at $t_\tau$. If $t_{\rm col} < t_\tau$ instead, the peak luminosity is defined by the merger-nova, and correspond to the maximum value of Eq.(\ref{eq:Lmn}). As a result, the luminosity function includes two components: one high-$L$ component related to wind dissipation, and another low-$L$ component related to merger-nova.

For GM1, it was found that $f_{\rm BH}:f_{\rm SMNS}:f_{\rm SNS}=40\%:30\%:30\%$ for prompt black holes, supra-massive NSs, and stable NSs \citep{gao16}. The latter two are relevant to the X-ray transients we model.

In order to determine the event rate density of these transients, we use sGRB event rate \citep{sun15} as a normalization. Since the event rate density evolves with redshift, throughout the paper we refer "event rate density" to the local one, i.e. $\rho_0$. It is possible that the sources other than NS-NS mergers (e.g. BH-NS mergers) may also contribute to the detected sGRBs, but in view of the prevalence of internal X-ray plateaus in sGRBs \citep{lv15}, it is possible that the majority of sGRBs are powered by NS-NS mergers. For simplicity, we assume that all sGRBs are powered by NS-NS mergers for the GM1 model. Based on the sGRB event rate density $\rho_{\rm sGRB}$, one can estimate that of the free-zone events:
\begin{equation}
\rho_{\rm free}=(f_{\rm SMNS}+f_{\rm SNS})k_{\Omega}\times \rho_{\rm sGRB}
\end{equation}
The event rate density of the trapped-zone events can be inferred based on the derived luminosity function from the simulations.

\begin{figure*}[hbtp]
\centering
\includegraphics[width=0.4\textwidth]{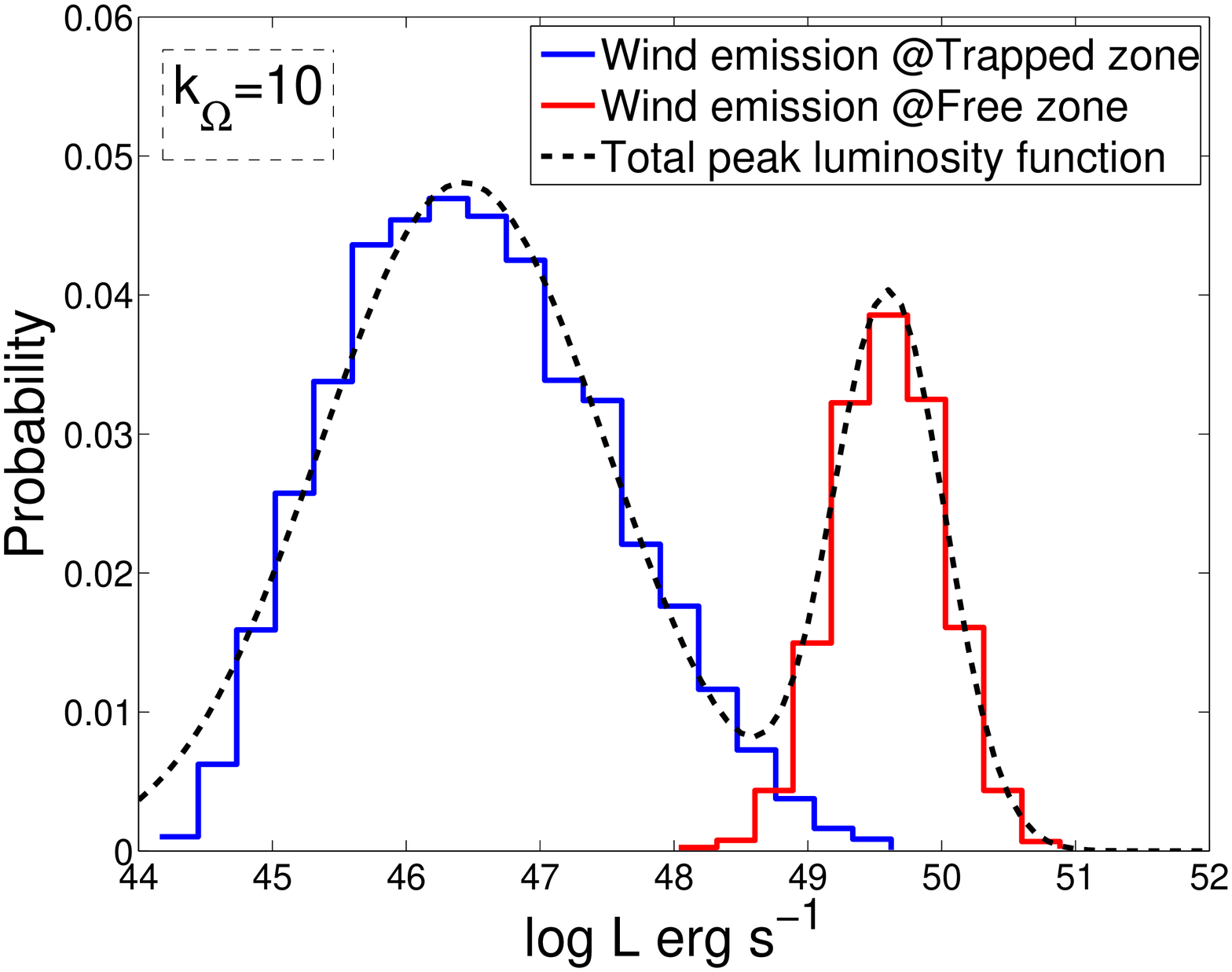}
\includegraphics[width=0.4\textwidth]{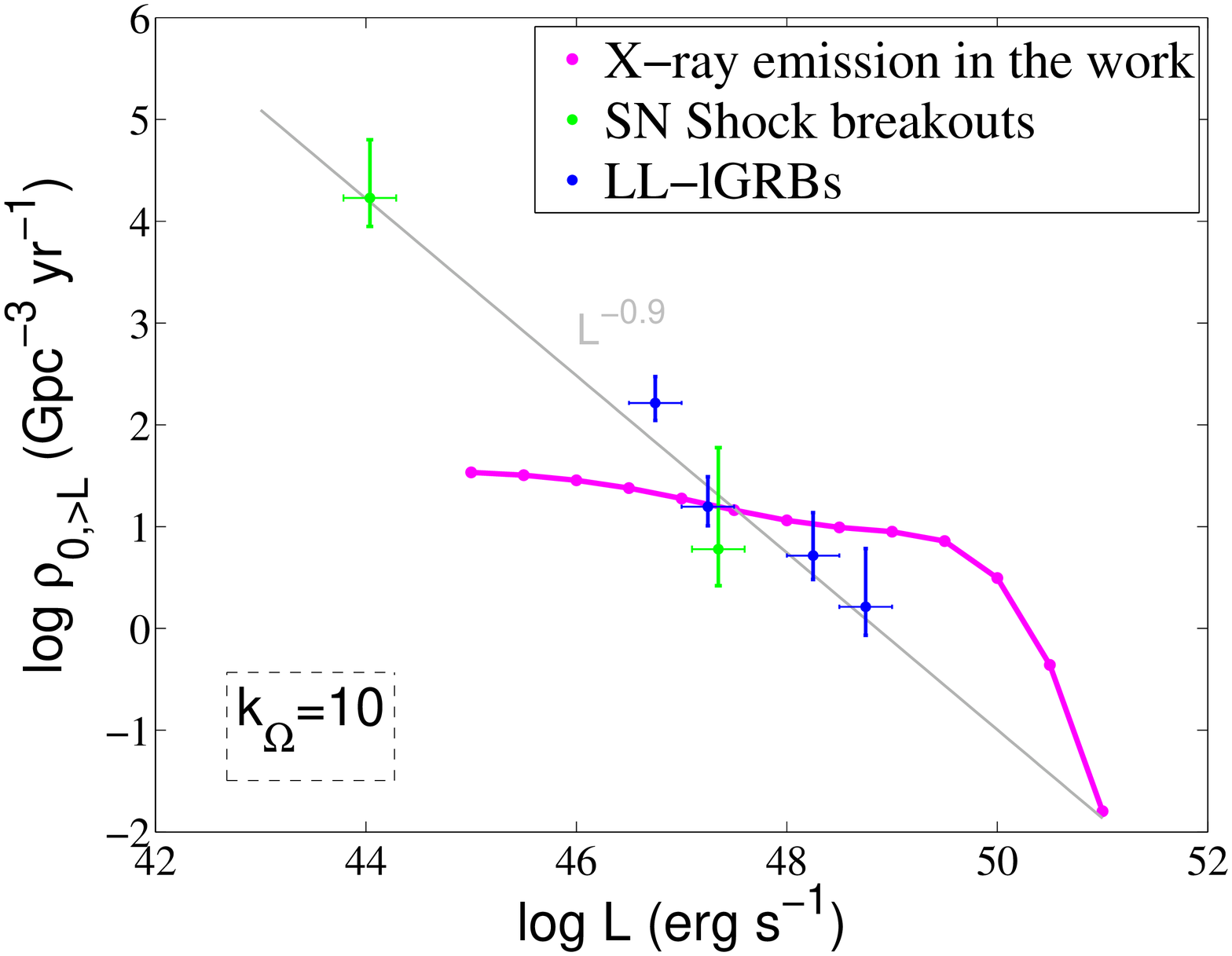}
\includegraphics[width=0.4\textwidth]{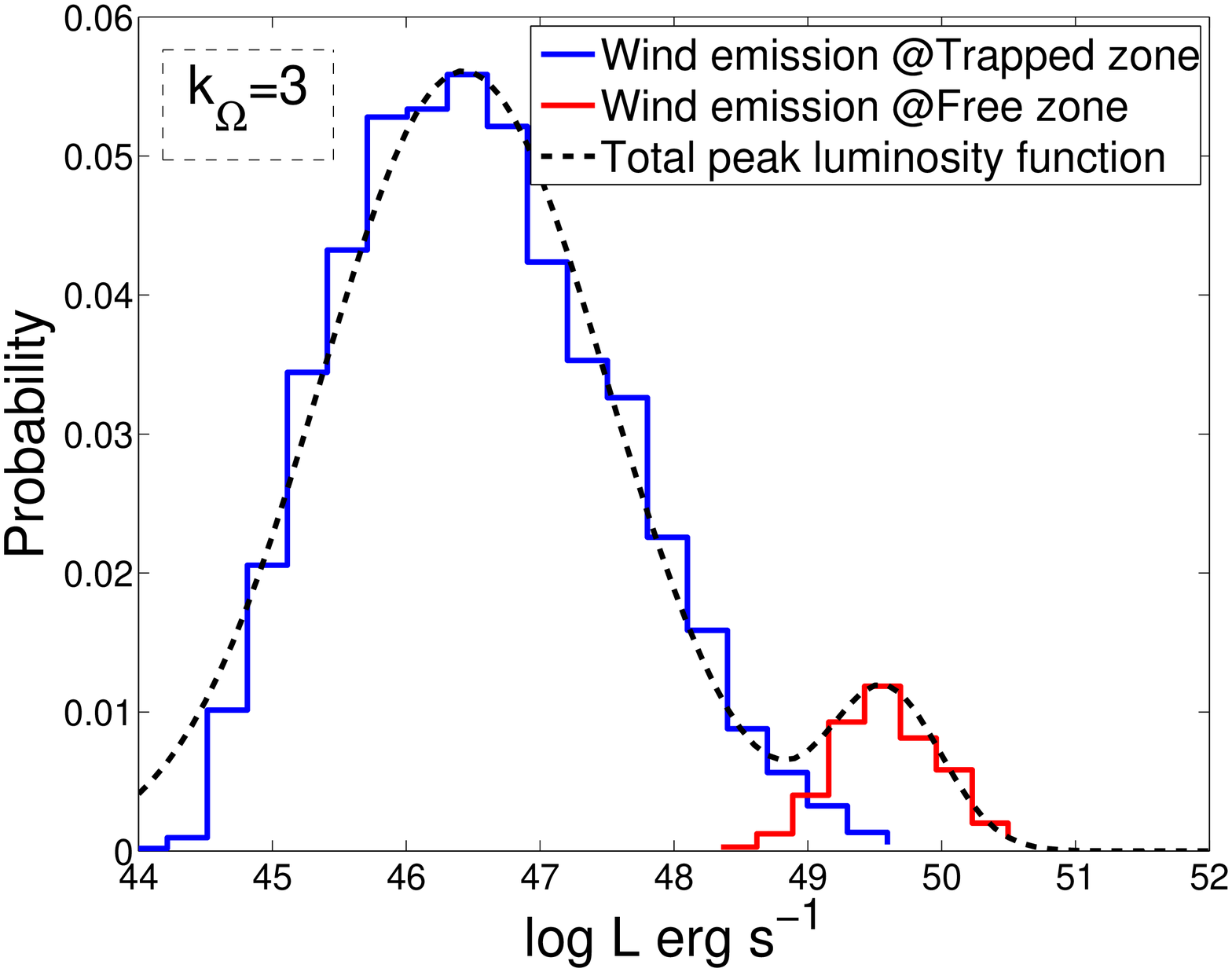}
\includegraphics[width=0.4\textwidth]{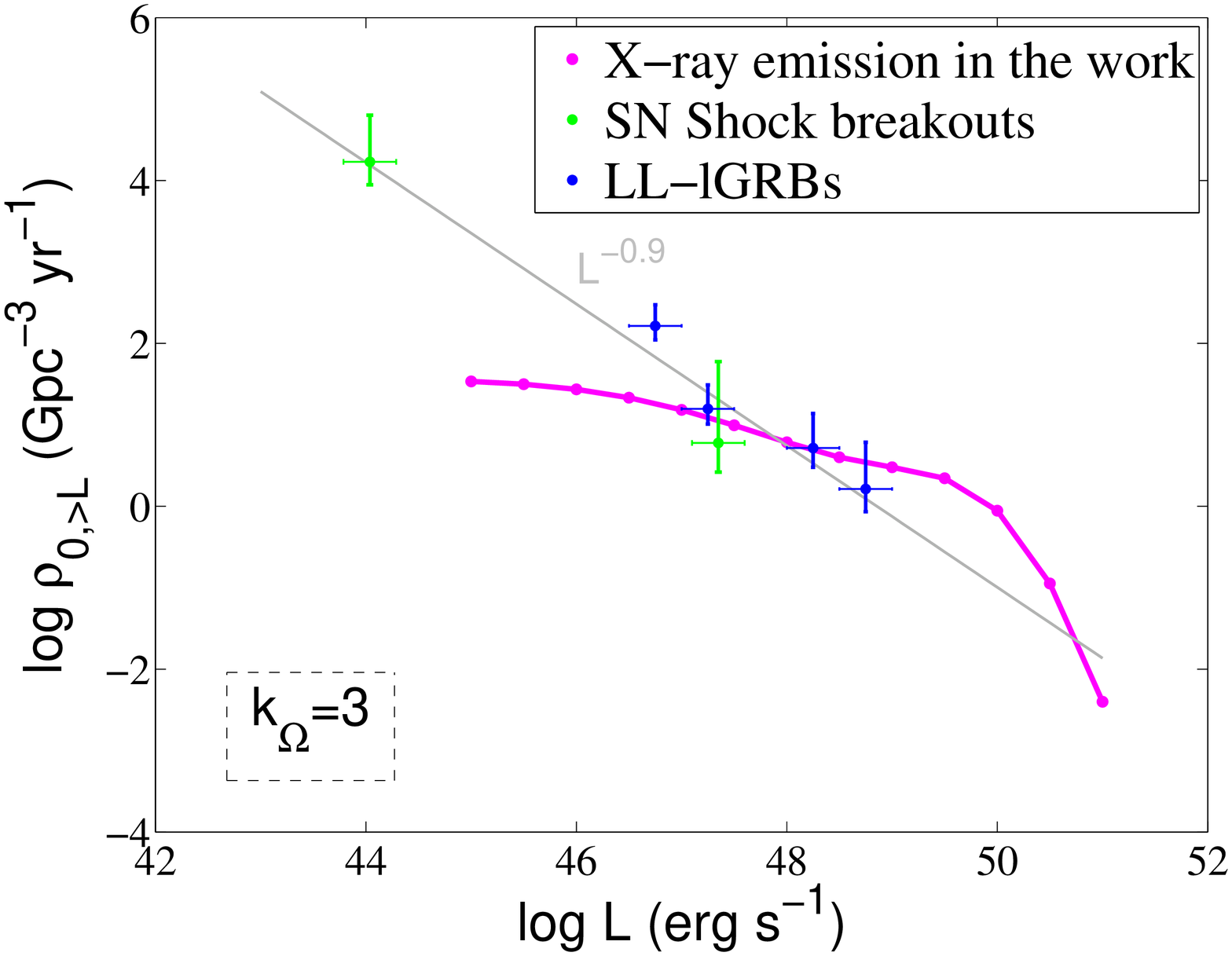}
\includegraphics[width=0.4\textwidth]{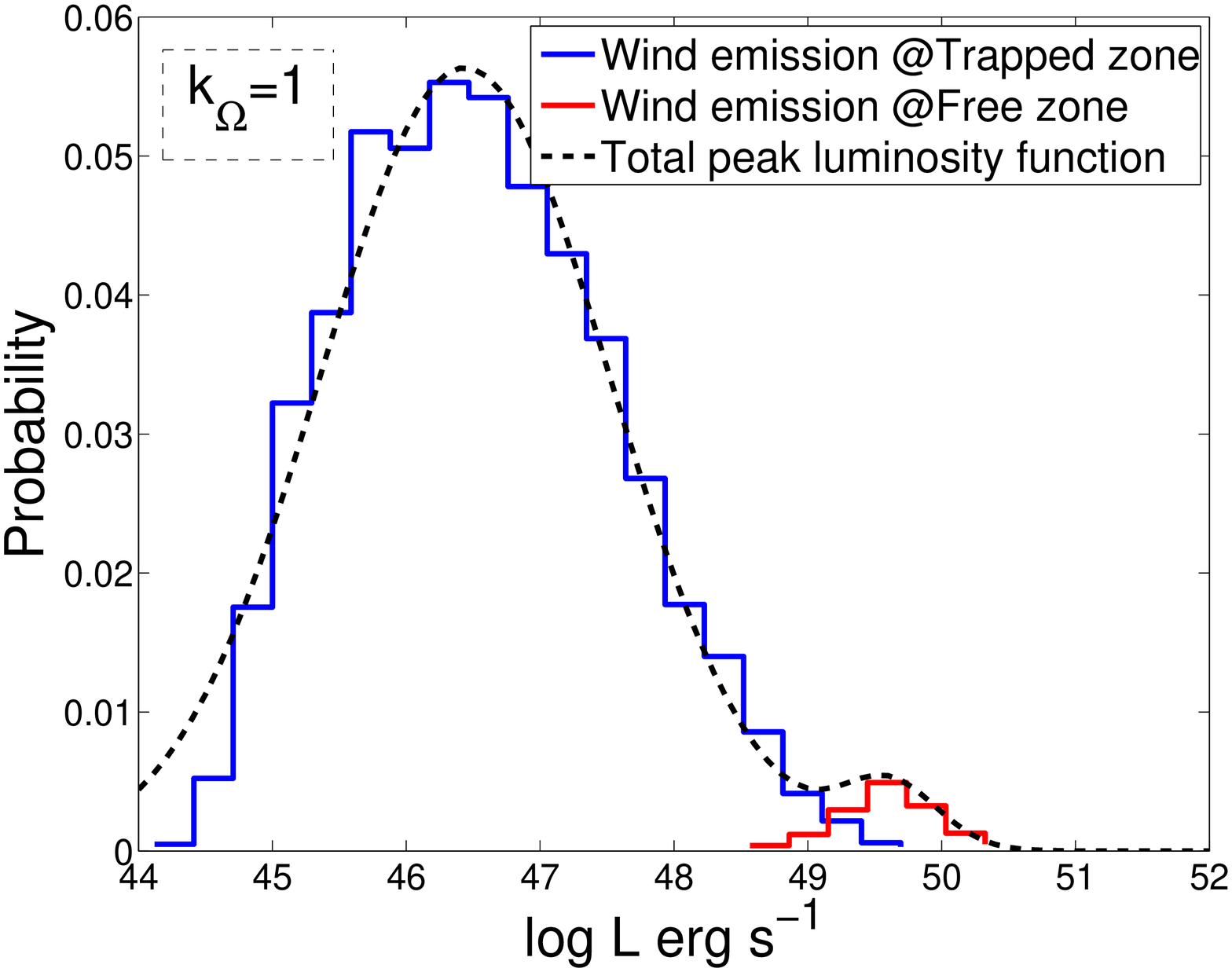}
\includegraphics[width=0.4\textwidth]{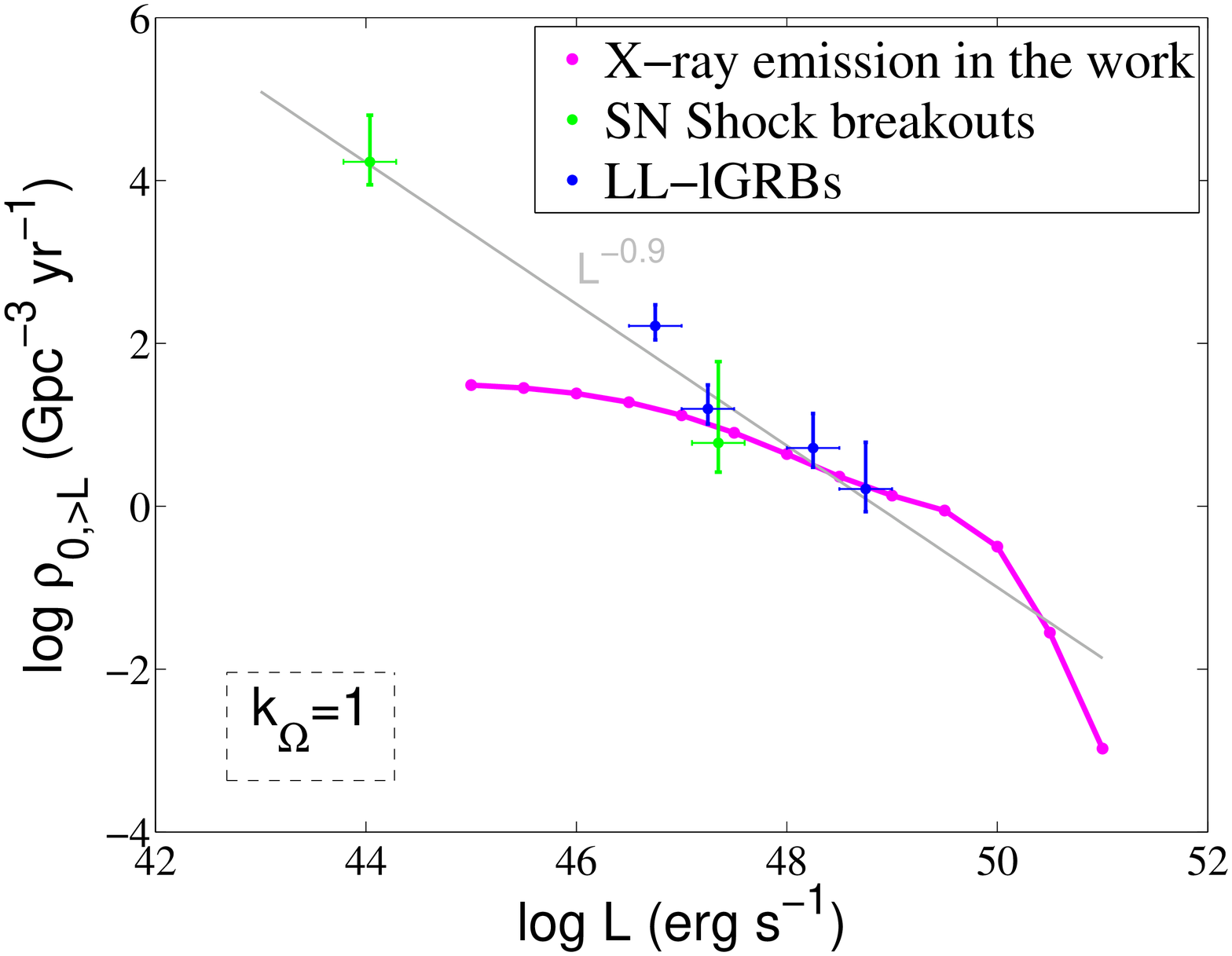}
\caption{Peak luminosity functions (left) and event rate densities (right) for the GM1 EoS for $k_{\Omega}=10,3,1$. Left: peak luminosity functions of both the free zone (red) and the trapped zone (blue). The X-ray transients from the merger-nova are so weak that they are neglected. The joint luminosity function is fit by the black dashed lines. Right: The event rate densities of the three models (pink) as compared with those of LL-lGRBs and SBOs \citep{sun15}. The joint fit of LL-lGRBs and SBOs and the powerlaw index is shown in grey.}
\label{fig:lf}
\end{figure*}

The luminosity functions and event rate densities of the X-ray transients for GM1 EoS are derived for three $k_\Omega$ values (10, 3, 1), which are shown in Figure \ref{fig:lf}. A typical jet opening angle of 10 degrees for sGRBs has been assumed. The peak luminosity functions are shown in the left column and the estimated event rate densities are shown in the right column. The peak LFs are bimodal and can be fit with the sum of two log-normal distributions 
\begin{eqnarray}
\Phi(d \log L)d\log L & = & \left[ N_{1}\exp\left(-\frac{(\log L-\mu_1)^2}{2\sigma_{1}^2}\right) \right.
\nonumber\\
&  + & \left. N_{2}\exp\left(-\frac{(\log L-\mu_2)^2}{2\sigma_{2}^2}\right) \right] d\log L.
\end{eqnarray}
The best-fit parameters are given in Table \ref{tab:lf}. 
The high-$L$ component due to the direct wind dissipation in the free zone (red) peaks at $10^{49.6}$ $\rm erg\,s^{-1}$, and the low-$L$ component from the trapped zone (blue) peaks at $10^{46.4}$ $\rm erg\,s^{-1}$. The high-$L$ component becomes progressively significant when $k_{\Omega}$ increases. In the event rate density plot, the high-$L$ component defines the plateau-like feature in the high-$L$ regime. For comparison, we display the event rate densities of SN shock breakouts and low-luminosity long GRBs (LL-lGRBs) from \cite{sun15}. All these three types of transients share the similar luminosity range from $10^{45}$ to $10^{50}$ $\rm erg\,s^{-1}$. Till now, we only have two confirmed SN SBOs and six LL-lGRBs. The non-detection of the X-ray transients discussed in this paper suggest that their event rate density should at most be comparable to the other two types.  This criterion places a strong constraint to $k_{\Omega}$. One can see that the case of $k_\Omega=10$ is already disfavored by the data. As a result, the free zone may be at most a few times of the jet zone. The majority of the solid angle should be in the trapped zone. This is consistent with the numerical simulations that suggest that the dynamical ejecta cover the majority of the solid angles. A more quantitative constraint on $k_\Omega$ may be carried out in the future when the X-ray / soft $\gamma$-ray transients are detected (e.g. Y. Li et al. 2016, in preparation, for a report on candidate events from the {\em Swift} BAT archives.)

\renewcommand\arraystretch{1.5}
\begin{table}[hbtp]
\centering
\caption{Best-fit parameters for the peak luminosity functions with two log-normal distributions for $k_{\Omega}=10, 3, 1$.}

\begin{tabular}{c|c|c|c|c|c}
\hline
\hline
Parameters & $\mu_{1}$ & $\sigma_{1}$ & $\mu_{2}$ & $\sigma_{2}$ & $N_{1}/N_{2}$ \\ 
\hline 
$k_{\Omega}=10$ & 46.4 & 1.1 & 49.6 & 0.4 & 1.2 \\ 
\hline 
$k_{\Omega}=3$& 46.4 & 1.1 & 49.6 & 0.4 & 4.5 \\ 
\hline 
$k_{\Omega}=1$& 46.5 & 1.1 & 49.6 & 0.4 & 13.6 \\ 
\hline 
\hline
\end{tabular}
\label{tab:lf} 
\end{table}

\subsection{Other EoSs and other high-energy transients}

Besides EoS GM1, we also test EoSs from \cite{li16c}, among which 6 out of 7 could satisfy the lower limit of the observed supra-massive NS fraction. On the other hand, the supra-massive NS fraction, $f_{\rm SMNS}$ for several EoSs (BSK21, Shen, CIDDM, CDDM1, CDDM2) are too large. In order to satisfy the observational constraints from sGRBs (which shows that $f_{\rm SMNS}$ is $\sim 30\% - 50\%$), for these EoSs, we assume that NS-NS mergers contribute to $50\%$ of the sGRB population, with the other half produced by NS-BH mergers.

\begin{figure}[hbtp]
\centering
\includegraphics[width=0.4\textwidth]{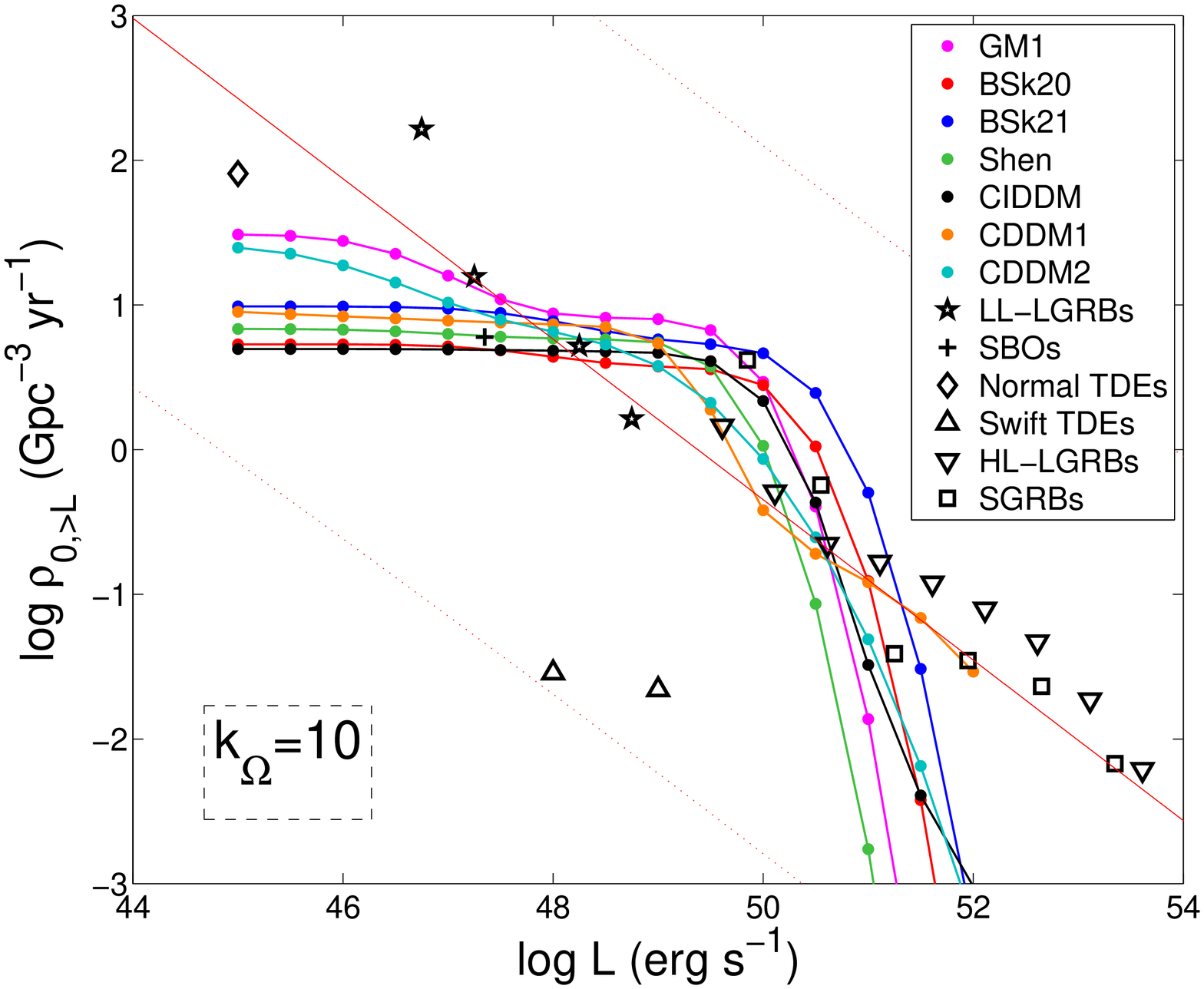} 
\includegraphics[width=0.4\textwidth]{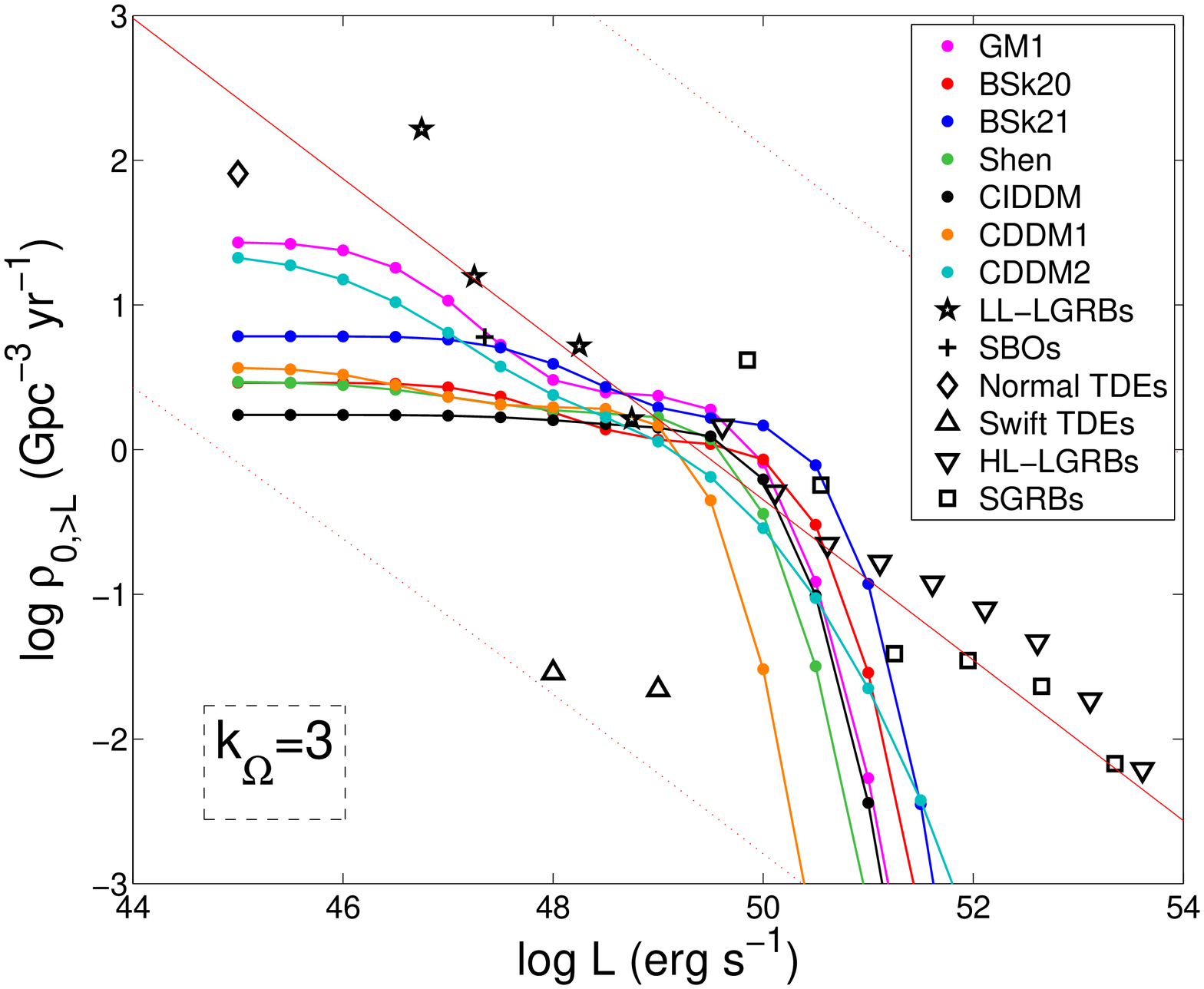} 
\includegraphics[width=0.4\textwidth]{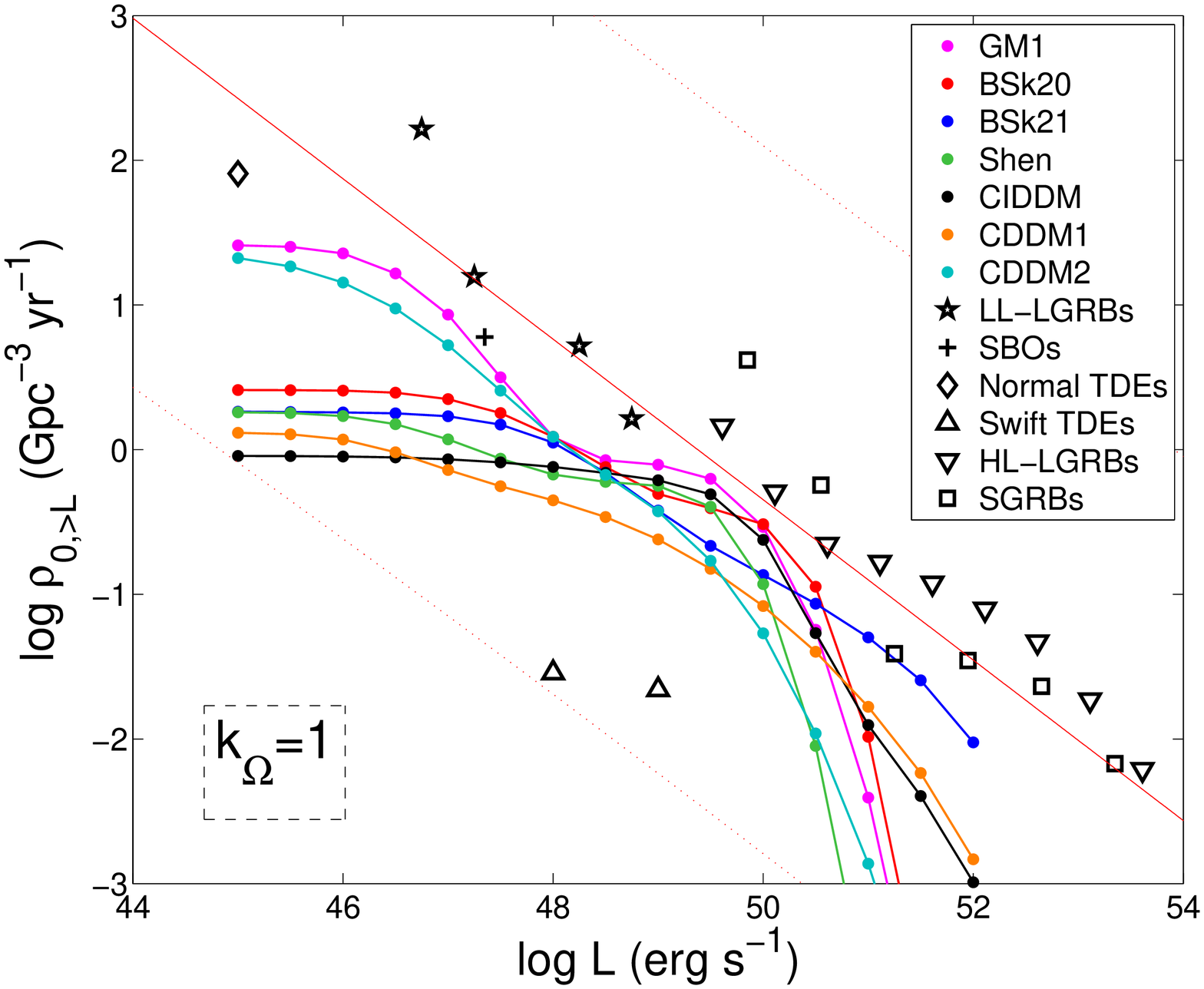}
\caption{Global event rate density distribution of the X-ray transients for $k_{\Omega}=10,3,1$ for seven EoSs studied in \cite{li16c}. The data of other high-energy transients, including LL-lGRBs, SBOs, TDEs and sGRBs, are also presented (from \cite{sun15}) with a single power law fit (red line) and $3\sigma$ boundary for the correlation (dotted line).}
\label{fig:EoSs}
\end{figure}

We show in Fig.\ref{fig:EoSs} the event rate density as a function of luminosity for all seven EoSs studied in \cite{li16c}.
The cases for $k_{\Omega}=10,3,1$ are shown in different panels. One can see that all EoSs except EoS BSk21 meets the non-detection criterion, i.e., the predicted X-ray transients have an event rate at most of other observed transients with a similar luminosity. With a larger solid angle ratio, say $k_{\Omega}\simeq 10$, all EoSs over-predict events with peak luminosity around $10^{49-50}$ $\rm erg\,s^{-1}$. The event rate density above $10^{45}$ $\rm erg\,s^{-1}$, $\rho_{0,>10^{45}}$, on the other hand, is around several tens of $\rm Gpc^{-3}\,yr^{-1}$ for EoSs GM1, CDDM2, but is one order of magnitude smaller for other EoSs. This is because this population of transients mostly depends on the fraction of stable compact stars, which guarantee to still produce X-ray emission when the ejecta become transparent. EoSs GM1 and CCDM2 (with 50\% correction for NS-NS mergers) have the highest stable fraction, and hence, the highest event rate density. Since at this luminosity the number is dominated by the trapped-zone low-$L$ component, $\rho_{0,>10^{45}}$ varies little for different $k_{\Omega}$ values.

Also plotted in Fig.\ref{fig:EoSs} are the event rate densities of all other high-energy transients (besides LL-lGRBs and SBOs, TDEs and sGRBs are also plotted) studied in \cite{sun15}. One can see that the predicted X-ray transients fall into the range of possible detections with the current facilities, if $k_\Omega$ is of order of unity.

\subsection{Detectability}

To be more quantitative, we calculate the detectability of such X-ray transients for present high-energy satellites and future wide-field X-ray telescopes based on the estimated luminosity functions and event rate densities. The detected number of events per year can be estimated as \citep{sun15}
\begin{equation}
\dot{N}=\frac{\Omega}{4\pi}\int_{L_{m}}^{L_{M}}\Phi(L)dL\int_{0}^{z_{max}(L)}\frac{\rho_0 f(z)}{1+z}\frac{dV}{dz}dz,
\end{equation}
where $\Phi(L)$ is the luminosity function and $f(z)$ describes the redshift distribution of events. We take $\Phi(L)$ for $k_{\Omega}=1$ for EoS GM1 as an example. The redshift distribution $f(z)$ is taken from Eq.(20) in \cite{sun15}, which considers a Gaussian distribution of the merger delay time scale for NS-NS mergers. The redshift-dependent specific co-moving volume reads (for the standard $\Lambda$CDM cosmology)
\begin{equation}
\frac{dV(z)}{dz}=\frac{c}{H_0}\frac{4\pi D_L^2}{(1+z)^2[\Omega_M(1+z)^3+\Omega_{\Lambda}]^{1/2}}.
\end{equation}
For a particular $L$, the maximum redshift $ z_{\rm max}(L)$, which defines the maximum volume inside which an event with luminosity $L$ can be detected, relies on the sensitivity threshold $F_{\rm th}$ via
\begin{equation}
F_{\rm th} = \frac{\eta L}{4\pi D_{L}^{2} (z_{\rm max})},
\end{equation}
We estimate the detection rates for {\em Swift}/BAT and XRT, {\em XMM-Newton}, {\em Chandra}, as well as the upcoming Chinese wide field X-ray telescope {\em Einstein Probe} (EP, with a designed field of view of 1str.).
We also consider the specific flux sensitivity for each telescope by assuming a $\sim 1000$s exposure time. 

\begin{figure*}[hbtp]
\centering
\includegraphics[width=0.45\textwidth]{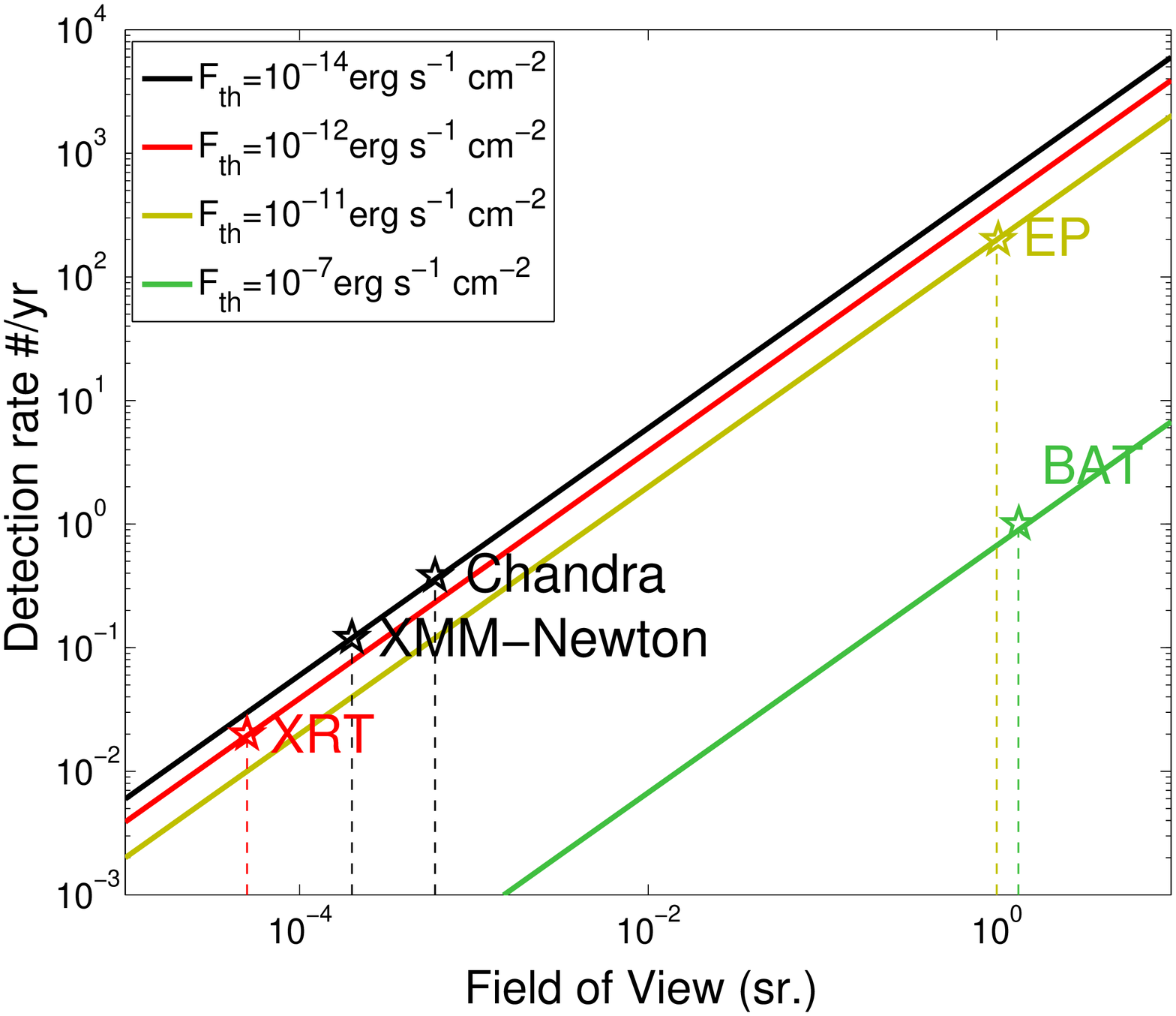} 
\includegraphics[width=0.45\textwidth]{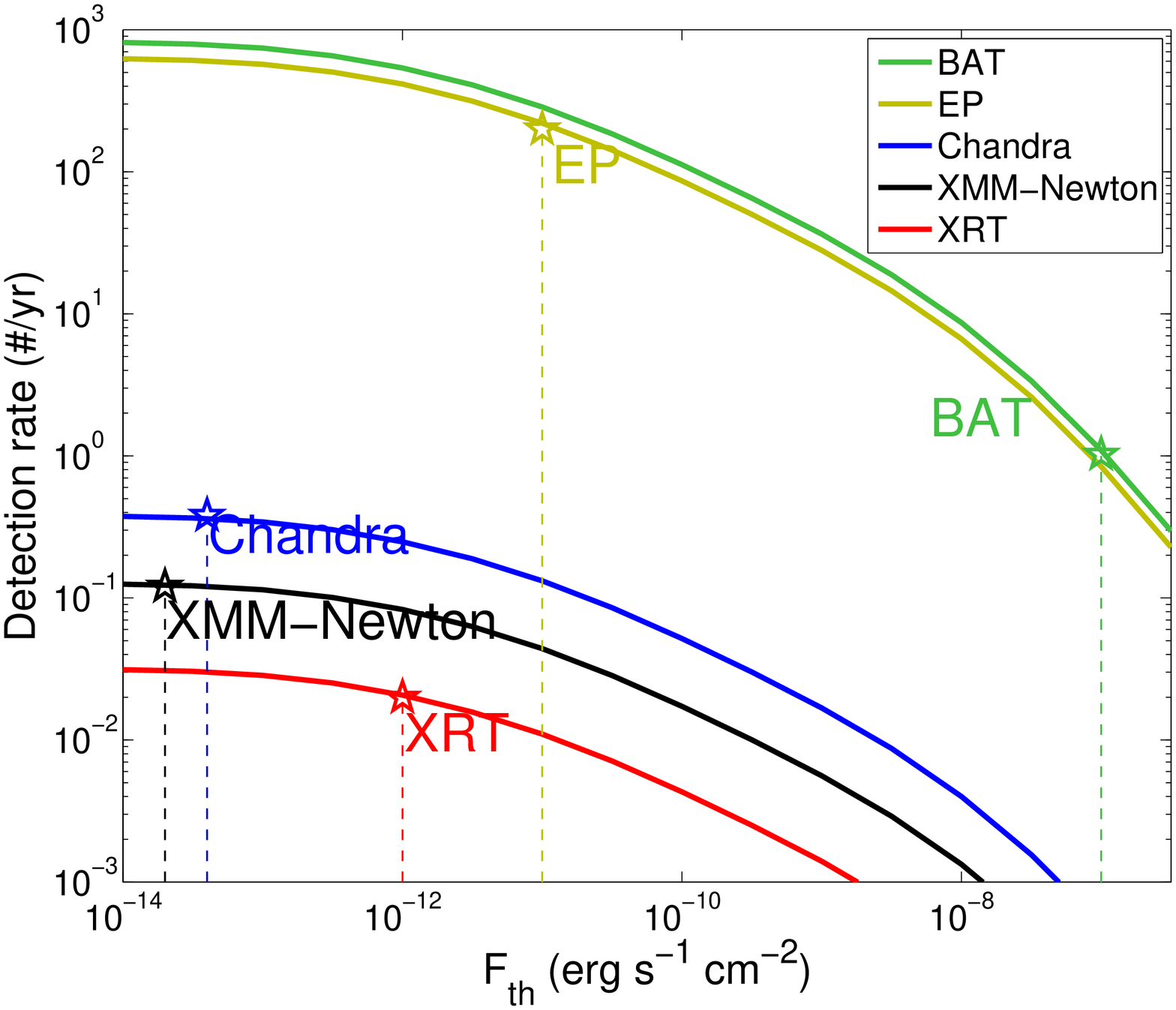} 
\caption{The detection rates of the proposed X-ray transients by the current and future high-energy detectors as a function of field of view (left) and sensitivity (right). Specific sensitivities of these instruments are considered and marked. We take $\Phi(L)$ for $k_{\Omega}=1$ for EoS GM1 as an example.}
\label{fig:dt}
\end{figure*}

In Figure \ref{fig:dt}, we give the detection rate as a function of both sensitivity and field of view for all the above-mentioned instruments. It can be seen that the present narrow field X-ray telescopes can hardly detect such X-ray transients. This is consistent with the non-detection of these events so far. The detection rate of BAT, $\dot{N}$, can be around 1-2 per year.  For ten years service, BAT may have already detected two dozens of such bright transients, if their spectra extend to the BAT energy band. However, they may have been confused as faint long-duration GRBs (or X-ray flashes). A systematic search in the faint BAT GRB sample may lead to identifications of such events (Y. Li et al. 2016, in preparation).

The prediction for EP is very promising. With the high sensitivity and large field of view, EP may be able to detect $\sim 100$ such events per year. Considering that for some parameters the light curves in the trapped zone show rapid decline as a function of time, which effectively reduce the integration time (with respect to the constant luminosity case), we more conservatively suggest a detection rate of several tens per solid angle per year for EP. The detections by EP would validate the existence of such events, testify the luminosity functions of the transients, and constrain the range of $k_\Omega$.

It is interesting to estimate the joint detection rate of these X-ray transients with the aLIGO GW signals. Taking EoS GM1 and $k_\Omega = 1$ as an example (the event rate densities vary by a factor of a few for different EoSs as discussed in section 3.3), within the 200 $\rm Mpc$ aLIGO average range for NS-NS mergers \citep{ligo}, the estimated event rate of these transients is about 1 per year all sky. The joint GW/X-ray detections depend on the field of view of the high-energy detector. Since 200 Mpc is so close, sensitivity would not be an important factor, and the field of view would be the key factor to determine the joint detection rate. For instruments like EP (about 1 steradian solid angle), one joint GW/X-ray detection may be made within a ten-year operation of both the X-ray telescope and the GW detectors. {\em Swift} BAT has a larger field of view ($\sim 1/7$ of all sky). It may give 1 joint detection with aLIGO in seven years.

\section{Conclusions $\&$ Discussion}

Considering that NS-NS mergers can leave behind a stable or a supra-massive millisecond magnetar, \cite{zhang13} proposed that there could be sGRB-less X-ray transients associated with gravitational wave events due to NS-NS mergers. In this paper, we studied such events in great detail. By defining three geometric zones (jet zone, free zone, and trapped zone) and solve the dynamical evolution and merger-nova ejecta, we predicted 12 different types of X-ray lightcurves for NS-NS mergers, 9 of which are sGRB-less. The X-ray transients are brighter from the free zone, with a typical luminosity $\sim 10^{49.6} \rm erg\,s^{-1}$. In the trapped zone, since X-ray luminosity rises only after the merger ejecta becomes transparent, the X-ray transients are fainter, with a typical luminosity of $\sim 10^{46.4} \rm erg\,s^{-1}$.

Through Monte Carlo simulations, we investigated the possible peak luminosity function and event rate density of the X-ray transients under different assumed NS/QS EoSs and for different assumed solid angle ratios, $k_\Omega$. In general, the peak luminosity function is bimodal, which can be fit with two log-normal distribution components from the free zone and the trapped zone, respectively. The relative number ratio between the two components depends on the unknown $k_\Omega$. By comparing the predicted event rate density of these transients with those of other known high-energy transients such as LL-lGRBs, sGRBs, TDEs, and tidal disruption events, we constrain that $k_\Omega$ is at most a few, which means that the free zone solid angle is at most comparable to (or slightly greater than) the sGRB solid angle. The event rate density of these transients above $10^{45}~{\rm erg~s^{-1}}$ is around a few tens of $\rm Gpc^{-3}\,yr^{-1}$ for EoSs GM1 and varies little for other EoSs. We calculate the detectability of these transients by current and future X-ray detectors. {\em Swift}/BAT may have detected some such transients, which might be confused as faint long GRBs or X-ray flashes. The upcoming sensitive, wide-field X-ray telescope such as the {\em Einstein Probe} mission may be able to detect up to several tens of events per year of such events. The joint aLIGO-high-energy detections should be rare, roughly 1 per year all sky. The detectability mostly depends on the field of view of the wide-field X-ray / soft $\gamma$-ray detectors.

\acknowledgments
We thank the referee for helpful comments and Phil Evans, Brian Metzger, and Daniel Siegel for useful discussion on the subjects of this paper. 
This work is partially supported by National Basic Research Program (973 Program) of China under Grant No. 2014CB845800. H.G. acknowledges support by theNational Natural Science Foundation of China under Grant No. 11543005, 11603003, 11633001, 11690024.

{}

\end{document}